\documentclass[a4paper]{article}
\usepackage[a4paper, total={6in, 9in}]{geometry}
\usepackage{floatrow}
\usepackage{hyperref}
\usepackage{filecontents}
\usepackage{subfigure}
\usepackage{bm}
\usepackage[font={scriptsize,bf}]{caption}
\usepackage{amsmath}
\usepackage{amsfonts}
\usepackage{amssymb}
\usepackage{graphicx}
\usepackage{bm}
\usepackage{multicol}
\usepackage{multirow}
\usepackage{epstopdf}
\usepackage{soul}
\usepackage{pifont}
\usepackage{lineno}
\usepackage[table]{xcolor}
\usepackage{color}
\usepackage[autostyle]{csquotes}
\usepackage{tikz,pgf}
\usepackage{collcell}

\usepackage{pgfplots}
\usepackage{wasysym}
\usepackage{lipsum}
\usepackage{algorithm} 
\usepackage{algpseudocode} 
\newlength\fheight
\newlength\fwidth
\newcommand{\cmmnt}[1]{\ignorespaces}
\usepackage{etoolbox}
\usepackage{setspace}
\usepackage{csquotes}

\title{Optimal Sparsity in Nonlinear Non-Parametric Reduced Order Models for Transonic Aeroelastic Systems}
\author{ \large Michael Candon$^1$\thanks{corresponding author, candon.michael@rmit.edu.au}, Errol Hale$^1$, Maciej Balajewicz$^2$, Arturo Delgado-Gutierrez$^1$ \\ \large and Pier Marzocca$^1$}

\date{
	\normalsize $^1$School of Engineering (Aerospace Engineering), RMIT University, Melbourne, AUS, 3000\\
	$^2$Independent Researcher, Boulder, CO, USA, 80304\\[2ex]%
}

\begin{document}
	
	\maketitle
	\begin{abstract}
        
        Machine learning and artificial intelligence algorithms typically require large amount of data for training. This means that for nonlinear aeroelastic applications, where small training budgets are driven by the high computational burden associated with generating data, usability of such methods has been limited to highly simplified aeroelastic systems. This paper presents a novel approach for the identification of optimized sparse higher-order polynomial-based aeroelastic reduced order models (ROM) to significantly reduce the amount of training data needed without sacrificing fidelity. Several sparsity promoting algorithms are considered, including; rigid sparsity, LASSO regression, and Orthogonal Matching Pursuit (OMP). The study demonstrates that through OMP, it is possible to efficiently identify optimized $s$-sparse nonlinear aerodynamic ROMs using only aerodynamic response information. This approach is exemplified in a three-dimensional aeroelastic stabilator model experiencing high amplitude freeplay-induced limit cycles. The comparison shows excellent agreement between the ROM and the full-order aeroelastic response, including the ability to generalize to new freeplay and velocity index values, with online computational savings of several orders of magnitude. The development of an Optimally Sparse ROM (OS-ROM) extends previous higher-order polynomial-based ROM approaches for feasible application to complex three-dimensional nonlinear aeroelastic problems, without incurring significant computational burdens or loss of accuracy.
        \end{abstract}

\section*{Nomenclature}

\begin{tabbing}
XXXXXXXXXX \= XX \= \kill
$\bm{d^p}$ \>\> Tensor of $p^{th}$-order Taylor partial derivatives\\
$\bm{D}$, $\bm{D_D}$, $\bm{D_L}$, $\bm{D_s}$ \>\> Flattened tensor of all Taylor partial derivatives for the full, diagonal, LASSO \\
\>\>and OMP derived ROMs\\
$\bm{\bar{D}}$, $\bm{\bar{D}_D}$, $\bm{\bar{D}_L}$, $\bm{\bar{D}_s}$, \>\> Tensor containing all $\bm{D}$, $\bm{D_D}$, $\bm{D_L}$, $\bm{D_s}$\\
$\bm{F_c}$ \>\> Nonlinear hinge restoring force vector\\
$\bm{F_v}$ \>\> Aerodynamic force vector (nodal coordinates)\\
$k$ \>\> Number of time lags\\
$k_\delta$ \>\> Freeplay axis rotational stiffness [Nm/rad]\\
$\bm{K_v}$ \>\> Stiffness matrix (nodal coordinates)\\
$\bm{K_B}$ \>\> Generalized stiffness matrix (fictitious mass coordinates)\\
$\bm{L}$ \>\> Lower triangular circulant matrix of inputs\\
$m$ \>\> Number of structural modes\\
$\bm{\mathcal{M}}$ \>\> Nonlinear matrix of inputs\\
$\bm{M_v}$ \>\> Mass matrix (nodal coordinates)\\
$\bm{M_B}$ \>\> Generalized mass matrix (fictitious mass coordinates)\\
$M_\delta$ \>\> Freeplay axis restoring moment\\
$M_\infty$ \>\> Freestream Mach number\\
$n$ \>\> Total number of samples\\
$n_s$ \>\> Total number of non-zero coefficients in $\bm{\bar{D}}$\\
$n_\kappa$ \>\> Total number of coefficients in $\bm{\bar{D}}$\\
$n$ \>\> Total number of samples\\
$N$ \>\> Number of structural nodes\\
$N_T$ \>\> Total number of tensors to be identified \\
$p$ \>\> Taylor polynomial order\\
$\bm{Q}$ \>\> Generalized aerodynamic force vector (fictitious mass coordinates)\\
$q_\infty$ \>\> Dynamic pressure [Pa]\\
$s$ \>\> Number of non-zero coefficients in $\bm{D_s}$\\
$\mathcal{T}$ \>\> Multi-variable Taylor series operator\\
$\bm{u}$ \>\> Displacement vector\\
$\bm{v}$, $\bm{\dot{v}}$, $\bm{\ddot{v}}$ \>\> Displacement, velocity and acceleration vector  (nodal coordinates)\\
$V^*$ \>\> Speed index\\
$\alpha$ \>\> $\ell_1$ regularization penalty term (LASSO)\\
$\alpha_0$ \>\> Freestream angle-of-attack [$^\circ$]\\
$\delta$ \>\> Freeplay axis rotation [$^\circ$]\\
$\delta_s$ \>\> Freeplay magnitude [$^\circ$]\\
$\kappa$ \>\> Total number of coefficients in $\bm{D}$\\
$\bm{\Phi_v}$ \>\> Matrix of normal modes\\
$\bm{\Phi_B}$ \>\> Matrix of baseline fictitious mass modes\\
$\omega$ \>\> Natural frequency\\
$\bm{\xi}$, $\bm{\dot{\xi}}$, $\bm{\ddot{\xi}}$ \>\> Generalized displacement, velocity and acceleration vectors\\

\end{tabbing}

\section*{Abbreviations}

\begin{tabbing}

XXXXXXXXXX \= XX \= \kill
AD \>\> Aerodynamic optimized\\
AE \>\> Aeroelastic optimized\\
AoA \>\> Angle-of-attack\\
CFD \>\> Computational fluid dynamics\\
DOF \>\> Degree-of-freedom\\
FOM \>\> Full order model\\
FM \>\> Fictitious masses\\
LASSO \>\> Least Absolute Shrinkage and Selection Operator\\
LCO \>\> Limit cycle oscillation\\
LS \>\> Least squares\\
OMP \>\> Orthogonal matching pursuit\\
PINV \>\> Moore-Penrose Pseudo-Inverse\\
ROM \>\> Reduced order model\\
D-ROM \>\> Diagonal sparsity reduced order model\\
L-ROM \>\> LASSO optimized reduced order model\\
OS-ROM \>\> Optimal sparsity reduced order model\\
SISO \>\> Single-input single-output\\
\end{tabbing}

	\section{Introduction}
	
    	High-performance aircraft are often required to perform high angle-of-attack (AoA) and abrupt maneuvering which leads to intense loading patterns on mechanical components. Partial (nonlinear hardening/softening or freeplay), or full failure of such components is highly problematic, leading to dangerous aeroelastic instabilities, such as, chaotic response or limit cycle oscillation (LCO).~\cite{conner97,lee99}
    
     	One of the challenges in modeling aeroelastic systems with discrete structural nonlinearities such as, freeplay, in the transonic flow regime, is the potential for nonlinear aerodynamic loads that coexist with the structural nonlinearity~\cite{candon17, candon19a}.  While linearization of the nonlinear structural model is common practice in aeroelastic problems involving freeplay (employing techniques such as fictitious masses~\cite{karpel75}), the assumption of aerodynamic linearization may, for some transonic systems, be invalid with small parameter changes--particularly pertinent at dynamic pressures close to the flutter boundary. If the nonlinear relationship between the structural displacement and the aerodynamic loading cannot be neglected, then the following options exist for the aeroelastician:
        \begin{itemize}
            \item Linearization of the nonlinear system can be performed with an accuracy penalization.
            \item The nonlinear forces on the structure can be solved for at every time interval with computational fluid dynamics (CFD) software which is computationally exhaustive. 
            \item Nonlinear aerodynamic reduced order models (ROM) can be used which are computationally efficient while retaining the fidelity of a full-order model (FOM). 
        \end{itemize}

        A ROM is a computationally efficient mathematical model of a complex physical system which retains only the features necessary for the practitioner to perform desired analysis. The range of ROM methodologies for transonic aerodynamic/aeroelastic applications are vast, including functional series~\cite{silva05}, proper orthogonal decomposition~\cite{hall00}, harmonic balance~\cite{thomas04a} and, more recently, dynamic mode decomposition~\cite{fonzi24, yao24} and deep neural networks~\cite{li19}. Dowell~\cite{dowell23} recounts his experience in model order reduction, including the history and state-of-the-art of the field.

        In the field of transonic aeroelasticity, often only the nonlinear aerodynamic forces on the elastic structure need to be realized, while information pertaining to the shock wave structure in the farfield can be disregarded. Accordingly, the functional series approach, most commonly a multi-variable Taylor series expansion (or Volterra series), in which nonlinear aerodynamic forces on the structure are described as a function of the structural response, is an intuitive choice. Other merits of this class of ROM include; $i$) relatively simplistic to implement and can be identified using standard system identification techniques, $ii$) the aerodynamic and structural response needed for identification can be obtained with minimal modification to standard CFD software, and $iii$) they intrinsically capture nonlinearity in their functional form. Despite significant advances in unsteady aerodynamic model reduction based on functional series in the past decades, application to three-dimensional aeroelastic systems has been limited to linearized or weakly nonlinear cases. Furthermore complex phenomena, such as, freeplay induced LCO is relatively scarce (for three-dimensional transonic aeroelastic systems with nonlinear aerodynamics).

        In this paper, the limitations that have prevented this class of ROM being applied to complex 3D aeroelastic systems are addressed through two pertinent ideas. Specifically, to identify transonic aerodynamic ROMs (of order greater than order 2) for three-dimensional aeroelastic systems (with several structural modes); $i$) a sparse formulation of the ROM is essential to minimize the amount of training data needed and therefore avoid the exhaustive offline computational burden that is associated with polynomial functionals and, $ii$) that automatic optimization of the sparsity patterns is necessary to efficiently obtain a robust model and avoid over fitting. Both of these points are comprehensively addressed by employing sparsity promoting algorithms, namely; LASSO regression and Orthogonal Matching Pursuit (OMP) to identify an optimal sparse representation of the first-, second- and third-order Taylor partial derivatives of the transonic aerodynamic forces, from input-output relations. Band-limited random excitation of each structural mode is used as an input to each mode and the corresponding generalized aerodynamic force vectors as outputs. The aerodynamic ROM is applied to a three-dimensional aeroelastic stabilator model with freeplay undergoing high amplitude LCO. Excellent correlation between the proposed ROM and full order aeroelastic model is observed. Out-of-sample performance is assessed by; $i$) optimizing the hyperparameters using aerodynamic cross-validation data only then applying the ROM to the aeroelastic problem, $ii$) applying the ROM to other unseen freeplay values, and $iii$) applying the ROM to unseen velocity index values. The novelty of this work is twofold:

        \begin{enumerate}
        \item A novel approach for the identification of optimized sparse nonlinear aeroelastic ROMs is presented which allows for a significant reduction in the amount of training data needed without sacrificing accuracy.
        \item This highly efficient method allows ROMs based on higher-order polynomial functionals to be applied to a three-dimensional aeroelastic structure that also contains structural nonlinearity for the first time. A task that would have been computationally exhaustive and/or inaccurate using previously proposed methods.
        \end{enumerate}

        The remainder of the paper is constructed as follows; in Section II the polynomial functional approximation of nonlinear unsteady aerodynamic forces is discussed and formulated. The proposed reduced order model with optimal sparsity is formulated in Section III. Section IV describes the nonlinear aeroelastic case study and practical details pertaining to implementation of the ROM. The results are presented and discussed in Section V. Section VI discusses the limitations of the proposed ROM and opportunities to extend the research. Section VII provided a final discussion and concluding remarks.

    \section{Higher-Order Taylor Series Expansion of the Unsteady Aerodynamic Forces} 

    \subsection{History of Unsteady Aerodynamic Reduced Order Models Based on Polynomial Expansions}
    
    Research that lays the foundation for the work presented in this paper considers unsteady aerodynamic ROMs based on Volterra theory. An important distinction to make is that the Volterra series is a generalization of the Taylor series or can be thought of as the Taylor series in multiple variables.
    
    The last 30 years has seen significant developments in transonic unsteady aerodynamic ROMs based on continuous- and discrete-time versions of Volterra theory, which includes applications in aeroelasticity. The implementation of Volterra theory is based on the identification of linear (aerodynamic impulse response) and nonlinear Volterra kernels.

    The earliest works in nonlinear aeroelasticity consider the identification of Volterra kernels by applying impulse/step functions (or variations of these) to the FOM and recording the linear/nonlinear aerodynamic impulse/step response~\cite{silva97, silva99,silva04,hong03,raveh01, marzocca04, marzocca05,balajewicz10}. 

    More recently it has been shown that for the identification of nonlinear kernels methods based on impulses are overly rigid and come with a range of limitations~\cite{balajewicz12, depaula15} (in particular when computational fluid dynamics codes are used to resolve the fluid loads), including:

    \begin{itemize}
		\item Numerical instability can occur when applying an impulse-type function to the system within a CFD solver. 
		\item Identifying kernels of order greater than two is computationally exhaustive, given that every component of the kernel requires a separate CFD simulation. 
		\item Ambiguity and a lack of control surrounds the selection of the magnitude and velocity (numerical time-step) of the impulses and the selection of these can require exhaustive parametric studies. 
	\end{itemize}

    An alternative approach is to excite the full-order model with a random signal of specific frequency and amplitude range, record the unsteady aerodynamic force response, then derive the linear and nonlinear kernels from the input-output relations using methods such as time-delay neural networks~\cite{depaula15, depaula19} or least squares(LS)~\cite{balajewicz09, balajewicz12}. In doing so, the limitations described above can be largely overcome, allowing for robust identification of higher-order kernels (up to $5^{th}$-order in these cases). In a very recent article, Brown~\cite{brown22} presents a multi-input Volterra-based approach, applied to a 3-DOF aeroelastic system. A blended step is used as excitation and $\ell_1$-regularized least-squares is used to derive the kernels which are expressed in terms of Laguerre polynomials to reduce the number of coefficients to be identified. The performance is generally very good. It is noted that out-of-sample performance in terms of reduced velocity is challenging in the nonlinear regime, as to be expected.

     \subsection{Taylor Series Expansion of the Unsteady Aerodynamic Forces}
     \label{sec:tsexp}
 
     Assuming that the unsteady aerodynamic forces on a structure can be described as a dynamic function of structural displacement according to

    \begin{equation}
    \label{eq:TS2}
        Q_n = f(\bm{u})
    \end{equation}

    \noindent where ${Q_n}$ represents the aerodynamic force at the current time interval, $\bm{u} = \{u_n, u_{n-1}, u_{n-2}, \hdots u_{n-k}\}^T$ represents displacements of a structural mode with the subscript denoting the discrete time interval, $k$ defines the number of time lags and $f()$ is an unknown dynamic function. Then, provided that the system is mildly nonlinear and memory fading $f(\bm{u})$ can be approximated using a multi-variable Taylor series expansion such that

    \begin{equation}
    \label{eq:TS3}
    f(\bm{u}) \approx \mathcal{T}(\bm{u})
    \end{equation}

   \noindent which can be evaluated at the location $\bm{u} = \bm{a}$ according to

   \begin{align}
    \begin{split}
    \label{eq:TS4}
    \mathcal{T}(u_n, \hdots , & u_{n-k}) = f(a_n, ..., a_{n-k}) + \sum^{n-k}_{j_1=n}\frac{\partial f(a_n, ..., a_{n-k})}{\partial u_{j_1}}(u_{j_1}-a_{j_1}) + \\ 
    &\frac{1}{2!}\sum^{n-k}_{j_1=n} \sum^{n-k}_{j_2=n}\frac{\partial^2 f(a_n, ..., a_{n-k})}{\partial u_{j_1} \partial u_{j_2}}(u_{j_1}-a_{j_1})(u_{j_2}-a_{j_2}) + \hdots \\
    + \frac{1}{p!}\sum^{n-k}_{j_1=n}\sum^{n-k}_{j_2=n} & \hdots \sum^{n-k}_{j_p=n} \frac{\partial^p f(a_n, ..., a_{n-k})}{\partial u_{j_1} \partial u_{j_2} \hdots \partial u_{j_p}}(u_{j_1}-a_{j_1})(u_{j_2}-a_{j_2}) \hdots (u_{j_p}-a_{j_p})
   \end{split}
   \end{align}
   
    \noindent where $p$ is the order of the Taylor expansion~\cite{duistermaat10}. This can be reduced to multi-index form and written as

    \begin{equation}
        \label{eq:TS5}
            \mathcal{T}(\bm{u}) = \sum_{|p| \geq 0} \frac{(\bm{u} - \bm{a})^p}{p!}\bm{d}^p
     \end{equation}

    \noindent where $\bm{d}^p = (\partial^p f)(\bm{a})$ is a $p^{th}$-order tensor that contains the $p^{th}$-order partial derivatives of $f(\bm{a})$, $i.e.$, $\bm{d}^1$ is the gradient of $f(\bm{a})$, $\bm{d}^2$ is the Hessian matrix, and so on. Given that $f(\bm{a})$ is not known a priori, the terms of the tensors $\bm{d}^0, \bm{d}^1, \hdots, \bm{d}^p$ are estimated from input-output training data, $i.e.$, the coefficients are identified to minimize the error between the Taylor approximation and the true values by $min || \bm{Q} - \mathcal{T}(\bm{u})||$. 

    \subsection{Motivation for Sparsity Promoting Algorithms}

    Identifying the coefficients of the Taylor partial derivatives is a linear problem~\cite{rugh81} which can be defined for a matrix of inputs $\bm{\mathcal{M}}$ and corresponding aerodynamic output vector $\bm{Q}$ (defined explicitly in the following section), given by
    
    \begin{equation}
    \label{eq:linprob}
        \bm{\mathcal{M} D} = \bm{Q} 
    \end{equation}
    
    \noindent where $\bm{D} = \{\bm{d}^0, \bm{d}^1, \hdots, \bm{d}^p \}$ contains the partial derivatives which are unknown.
    
    Although any standard least squares (LS) approach can be used to solve the inverse problem for $\bm{D}$, the number of coefficients in $\bm{D}$ to be estimated grows exponentially with the order of the Taylor expansion and the number of time lags being considered - known as the curse of dimensionality. This means that the number of samples in $\bm{Q}$ required to identify the coefficients also grows exponentially. This can quickly become computationally exhaustive when considering that $i$) training data is generated with a CFD code, $ii$) complex 3D aeroelastic models typically have a large number of cells in the fluid mesh, and $iii$) complex 3D aeroelastic models also have a large number of structural modes. 

    In the work of Balajewicz and Dowell~\cite{balajewicz12}, it is proposed that this curse of dimensionality can be avoided by identifying a sparse representation of $\bm{{D}}$. It is shown that by neglecting the nonlinear lag terms the total number of coefficients to be estimated becomes a linear function of the polynomial order. Although estimation of the main diagonal is an intuitive choice given the memory-fading nature of transonic aeroelastic systems, it cannot be considered as a general rule. There may be systems that require some of the lag terms to be identified, in which case any attempt to identify an optimal sparsity pattern becomes intractable.  

    With these points in mind, it is clear that for feasible application of this class of ROM to complex 3D aeroelastic problems; $i$) a sparse representation of the Taylor derivatives is required to minimize the amount of training data needed, and $ii$) automated sparsity pattern selection is necessary to facilitate rapid ROM generation for large models and avoid over-fitting~\cite{candon24a}. In solving the inverse linear problem defined in Eq.~\ref{eq:linprob}, this paper employs three techniques to promote sparsity, namely; $i$) \textbf{Rigid Sparsity~\cite{balajewicz12}} (as a benchmark) where the sparsity pattern is embedded in the matrix of inputs, $ii$) \textbf{Orthogonal Matching Pursuit~\cite{foucart13}} a greedy algorithm that recovers a sparse representation of a signal in a step-by-step iterative manner, and $iii$) \textbf{LASSO Regression~\cite{tibshirani96}} that uses $\ell_1$ regularization as a penalty term in solving the least squares optimization problem. These three sparsity promotion techniques are discussed in further detail in the following section.

    \section{Nonlinear Aeroelastic Reduced Order Model with Optimal Sparsity}
    \label{sec:roms}

    In this section, the procedure for identifying the aeroelastic ROMs is described. This definition is for the single-input single-output (SISO) variant of the identification procedure, $i.e.$, nonlinear interactions between structural modes are neglected. Although the multi-input identification procedure has been shown to provide superior performance~\cite{balajewicz09}, it is a secondary avenue for the curse of dimensionality. Specifically, the number of tensors (or kernels) $N_T$ to be identified grows exponentially as a function of the number of structural modes $m$ and the order $p$, according to $N_T = m^p$. So, from a practical perspective for 3D aeroelastic problems, neglecting the nonlinear cross-terms can be thought of as a further means of promoting sparsity which, if possible, is desired. A more detailed discussion on this is provided in Section~\ref{sec:disc}.

    \subsection{Band Limited Random Excitation}
    
    The first step in creating any unsteady aerodynamic ROM is to perturb the structural modes within a full-order aerodynamic solver and to record the aerodynamic responses. In this approach, the structural model is excited using band limited random noise within a finite volume CFD code. Using the SISO identification procedure, each structural mode is excited in isolation. The amplitude and frequency band of the excitation functions are chosen by estimating the LCO amplitudes and frequencies in the aeroelastic response. 
    
    An important consideration is to observe a smooth transition from the undeformed structure and converged steady-state fluid forces. The reason being that any discontinuity ($i.e.$, a step-like change in displacement) will cause spurious aerodynamic response information and inaccuracies in the ROM identification process. A hyperbolic tangent is applied to the the raw signal ${\bm{u}}_{j,r}$, ensuring a smooth transition to the modal excitation over the first 20 time intervals, given for $n$ total input samples by

        \begin{equation}
        \label{eq:smramp}
            \bm{u}_j = \left(0.5\tanh{\left( \frac{n-20}{2} \right)+1}\right)\bm{u}_{j,r}
        \end{equation}

    \noindent where $\bm{u}_j$ is the band limited random signal used to excite the $j^{th}$ structural mode. 
    
    \subsection{Input and Output Matrices}

     Considering a total of $m$ structural modes, the vector of $n$ output training samples $\bm{Q}^{ij}$ is obtained using the full-order aerodynamic model to excite each $j^{th}$ structural mode individually with $\bm{u}_j \in \mathbb{R}^n, j = 1\hdots m$. The forces are then projected onto each $i^{th}$ structural mode to give the generalized aerodynamic force vector
     
     \begin{equation}
         \bm{Q}^{ij} = \{Q_1^{ij},Q_2^{ij}, \hdots , Q_n^{ij}\}^T \in \mathbb{R}^n, i = 1\hdots m, j = 1\hdots m.
     \end{equation} 
     
     To construct the matrix of inputs, first a lower left triangular circulant matrix is constructed from $\bm{u}_j$ (truncated for $k$ time lags) to give
     \begin{align} 
        \label{eq:TS8}
        \bm{L}^j = \begin{bmatrix}
        u_{j_1} & 0 & 0 & \hdots & 0\\
        u_{j_2}  & u_{j_1} & 0 & \hdots &0 \\
        \vdots  &\vdots & \vdots& \ddots & \vdots \\
        u_{j_n}  & u_{j_{n-1}}  & u_{j_{n-2}}  &\hdots & u_{j_{n-k}}\\
      \end{bmatrix}\in \mathbb{R}^{n \times k}, j = 1\hdots m
      \end{align}
     \noindent and the $p^{th}$-order multivariable Taylor expansion of the rows gives 

     \begin{equation}
             \bm{\mathcal{M}}^j = \mathcal{T}(\bm{L}_{n*}^j)  = [\bm{\mathcal{M}}_1^j, {\bm{\mathcal{M}}}_2^j, \hdots, {\bm{\mathcal{M}}}_p^j] \in \mathbb{R}^{n \times \kappa}, j = 1\hdots m
     \end{equation}

     \noindent where $\kappa = \sum_{p_i=1}^{p} \binom{k+(p_i-1)}{p_i}$ and $\bm{\mathcal{M}}^j$ contains all monomials of $\bm{L}^j$ up to order $p$, defined explicitly as

    \begin{equation}
    \bm{\mathcal{M}}_1^j = \bm{L}^j
    \end{equation}
    \begin{align} 
        \label{eq:TS8}
        {\bm{\mathcal{M}}}_2^j = \begin{bmatrix}
        u_{j_1}u_{j_1} & 0 & 0 & \hdots & 0\\
        u_{j_2}u_{j_2}  & u_{j_2}u_{j_1} & u_{j_2}u_{j_2} & \hdots &0 \\
        \vdots  &\vdots & \vdots& \ddots & \vdots \\
        u_{j_n}u_{j_n}  & u_{j_n}u_{j_{n-1}} & u_{j_n}u_{j_{n-2}}  &\hdots & u_{j_{n-k}}u_{j_{n-k}}\\
      \end{bmatrix}
      \end{align}

     \begin{align} 
        \label{eq:TS8}
        {\bm{\mathcal{M}}}_p^j = \begin{bmatrix}
        \prod_{p_i=1}^p\{u_{j_1}\} & 0 & 0 & \hdots & 0\\
        \prod_{p_i=1}^p\{u_{j_2}\}  & \prod_{p_i=1}^{p-1}\{u_{j_2}\} \cdot u_{j_1} & \prod_{p_i=1}^{p-2}\{u_{j_2}\} \cdot \prod_{p_i=1}^{2}\{u_{j_1}\} & \hdots &0 \\
        \vdots  &\vdots & \vdots& \ddots & \vdots \\
        \prod_{p_i=1}^p\{u_{j_n}\}  & \prod_{p_i=1}^{p-1}\{u_{j_n}\} \cdot u_{j_{n-1}}  & \prod_{p_i=1}^{p-1}\{u_{j_n}\} \cdot u_{j_{n-2}}  &\hdots & \prod_{p_i=1}^{p}\{u_{j_{n-k}}\}\\
      \end{bmatrix}
      \end{align}
      
     The sparsity can also be predefined and embedded in the matrix of inputs which is referred to as rigid in this work as it is an alternative to optimal coefficient selection. The matrix of inputs for the identification of the diagonal terms only~\cite{balajewicz12} (diagonal sparsity) is defined as

     \begin{equation}
        \bm{\mathcal{D}}^{j} = [\bm{L}^j, \bm{L}^{j^2}, \hdots, \bm{L}^{j^p}] \in \mathbb{R}^{n \times pk}, j = 1\hdots m
     \end{equation}

    \subsection{Identification of the Reduced Order Model using Least Squares}
    
    The partial derivatives without sparsity can be identified using least squares, evaluated according to
     
    \begin{equation}
        \bm{D}^{ij} = {\bm{\mathcal{M}}^{j}}^+\bm{Q}^{ij}  \in \mathbb{R}^{\kappa}, i = 1\hdots m, j = 1\hdots m
    \end{equation}

    \noindent where $^+$ is the Moore-Penrose Pseudo-Inverse (PINV) and $\bm{D}^{ij}$ contains the flattened tensors of partial derivatives corresponding to the generalized aerodynamic forces for $\bm{Q}_i(\bm{u}_j)$. The final ROM is obtained by iterating through all combinations of $\bm{\mathcal{M}}^j$ and $\bm{Q}^{ij}$ where all instances of $\bm{D}^{ij}$ are stored in a three-dimensional array $\bm{\bar{D}} \in \mathbb{R}^{m \times m \times \kappa}$.


    Similarly, the set of partial derivatives with diagonal sparsity can be identified using least squares, evaluated according to

    \begin{equation}
        \bm{D_D}^{ij} = {\bm{\mathcal{D}}^j}^+\bm{Q}^{ij}  \in \mathbb{R}^{pk}, i = 1,\hdots, m, j = 1\hdots m
    \end{equation}

    \noindent where $\bm{D_D}^{{ij}}$ contains the diagonals of the derivative tensors corresponding to the generalized aerodynamic forces for $\bm{Q}_i(\bm{u}_j)$. The final \textbf{D}iagonal sparsity \textbf{ROM} (D-ROM) is obtained by iterating through all combinations of $\bm{\mathcal{D}}^j$ and $\bm{Q}^{ij}$ where all instances of $\bm{D_D}^{ij}$ are stored in a three-dimensional array $\bm{\bar{D}_D} \in \mathbb{R}^{m \times m \times kp}$.

    
    \subsection{Identification of the Reduced Order Model using Orthogonal Matching Pursuit}
    \label{sec:OSROM}
    
    In this section the OMP-based formulation of the nonlinear unsteady aerodynamic ROMs with optimal sparsity (OS-ROM) is defined.~\cite{foucart13} Considering the linear problem described in Eq.~\ref{eq:linprob}, the objective of the OMP-based identification strategy is to identify a sparse representation of $\bm{D}$, denoted by $\bm{D_s}$, by adding terms to $\bm{D_s}$ iteratively until a pre-defined stopping criterion is met. This requires the following $\ell_0$-minimization problem to be solved

    \begin{equation*}
    \label{eq:TS16}
        \underset{\bm{D_s}}{\operatorname{argmin}}  ||\bm{D_s}||_0 \quad \textrm{subject to} \quad \bm{\mathcal{M}\bm{D_s}} = \bm{Q}
    \end{equation*}

     \noindent where $||\bm{D_s}||_0$ is the $\ell_0$ pseudo-norm or the number of non-zero elements in $\bm{D_s}$. Assuming that $\bm{D_s}$ is $s$-sparse ($s_{D_s} \geq ||\bm{D_s}||_0$), it can be recovered exactly by OMP if $\bm{\mathcal{M}}$ and $\bm{D_s}$ satisfy following inequality:

    \begin{equation}
    \label{eq:TS17}
    \mu_\mathcal{M} < \frac{1}{2s_{D_s} - 1}
    \end{equation}

     \noindent where $\mu_\mathcal{M}$ is the mutual coherence of the columns of $\bm{\mathcal{M}}$ and $s_{D_s}$ is the sparsity of $\bm{D_s}$. From Eq.~\ref{eq:TS17}, $\bm{D_s}$ can be at most $\frac{1}{2\mu_\mathcal{M}}$-sparse. Using the number of non-zero terms as the stopping criterion, the coefficients in $\bm{D_s}$ are identified in a step-by-step iterative manner using the OMP Algorithm 1 as follows

    \begin{algorithm}
	\caption{OMP ($\bm{\mathcal{M}}$, $\bm{Q}$)}
        \begin{flushleft}
         \hspace*{\algorithmicindent} \small{\textbf{Input:} $\bm{\mathcal{M}}$, $\bm{Q}$} \\
         \hspace*{\algorithmicindent} \small{\textbf{Result:}  ${\bm{D_s}}_k$}
         \end{flushleft}
	   \begin{algorithmic}[1]
                \label{alg:omp}
                \State Initialization $\bm{r_0} = \bm{Q}$, $\Lambda_0 = \varnothing$;
                \State Normalize all columns of $\bm{\mathcal{M}}$ to unit $L_2$ norm (optional);
                \For {$k=1,2,\hdots$}
                    \State $\lambda_k = \underset{j \notin \lambda_{k-1}}{\operatorname{argmax}} | \langle \bm{a_j},\bm{r_{k-1}}\rangle |$
                    \State  $\lambda_k =\lambda_{k-1} \cup {\lambda_k}$
                    \State ${\bm{D_s}}_k (i \in  \lambda_k) = \underset{\bm{D_s}}{\operatorname{argmin}}||\bm{\mathcal{M}}_{\lambda_k}\bm{D_s}-\bm{Q}||_2, \quad {\bm{D_s}}_k (i\notin \lambda_k) = 0$
                    \State $\bm{\hat{Q}}_k = \bm{\mathcal{M}} {\bm{D_s}}_k$
                    \State $\bm{r}_k \leftarrow \bm{Q} - \bm{\hat{Q}}_k$
                \EndFor
	\end{algorithmic} 
    \end{algorithm}

    The partial derivatives for the \textbf{O}ptimal \textbf{S}parsity \textbf{ROM} (OS-ROM) can be identified for $\bm{\mathcal{M}}^j$ and $\bm{Q}^{ij}$ using the OMP Algorithm 1 according to
    
     \begin{equation}
    \label{eq:OMPsol}
    \bm{D_s}^{ij} = \mathrm{OMP}(\bm{\mathcal{M}}^j,\bm{Q}^{ij})  \in \mathbb{R}^{\kappa}, i = 1\hdots m, j = 1\hdots m
    \end{equation}
    
    
    

    \noindent where $||\bm{D_s}^{ij}||_0 << ||\bm{D}^{ij}||_0$. Finally, iterating through each structural mode, all instances of $\bm{D_s}^{ij}$ are stored in a three-dimensional array $\bm{\bar{D}_s} \in \mathbb{R}^{m \times m \times \kappa}$. 


    \subsection{Identification of the Reduced Order Model using LASSO Regression}
    \label{sec:OSROM}

    An alternative to OMP is to solve the optimization problem with sparsity-inducing regularizers. Perhaps best known is the Least Absolute Shrinkage and Selection Operator (LASSO)~\cite{tibshirani96} algorithm which uses $\ell_1$ regularization as a penalty term in the least squares solution of the inverse linear problem (Eq.~\ref{eq:linprob}), therefore recovering an optimal sparse representation of the coefficients. For this the objective is to solve the $\ell_1$-minimization problem

    \begin{equation*}
    \label{eq:LAS2}
        \mathrm{argmin}  ||\bm{\mathcal{M} D_L}-\bm{Q}||_{2}^{2} + \alpha||\bm{D_L}||_1
    \end{equation*}

    \noindent where $||\bm{D_L}||_1 = \sum_{k}^{n}|{D_L}_k|$ denotes the $\ell_1$-norm of $\bm{D_L}$ and $\alpha > 0$ is the regularization parameter. The set of partial derivatives for the \textbf{L}ASSO optimized \textbf{ROM} (L-ROM), can be identified by solving the $\ell_1$-minimization problem for $\bm{\mathcal{M}}^j$ and $\bm{Q}^{ij}$, identifying $\bm{D_L}^{ij} \in \mathbb{R}^{\kappa}, i = 1\hdots m, j = 1\hdots m$ which contains the flattened tensors of sparse partial derivatives corresponding to the generalized aerodynamic forces for $\bm{Q}_i(\bm{u}_j)$. Finally, iterating through each structural mode, all instances of $\bm{D_L}^{ij}$ are stored in a three-dimensional array $\bm{\bar{D}_L} \in \mathbb{R}^{m \times m \times \kappa}$. 


	\section{Nonlinear Aeroelastic Framework}
	
	\subsection{Modified AGARD 445.6 Wing}
	\label{sec:model} 
	
\begin{figure}[h]
		\centering
		\subfigure[]{\label{agardBENCH}
			\includegraphics[width=0.45\textwidth]{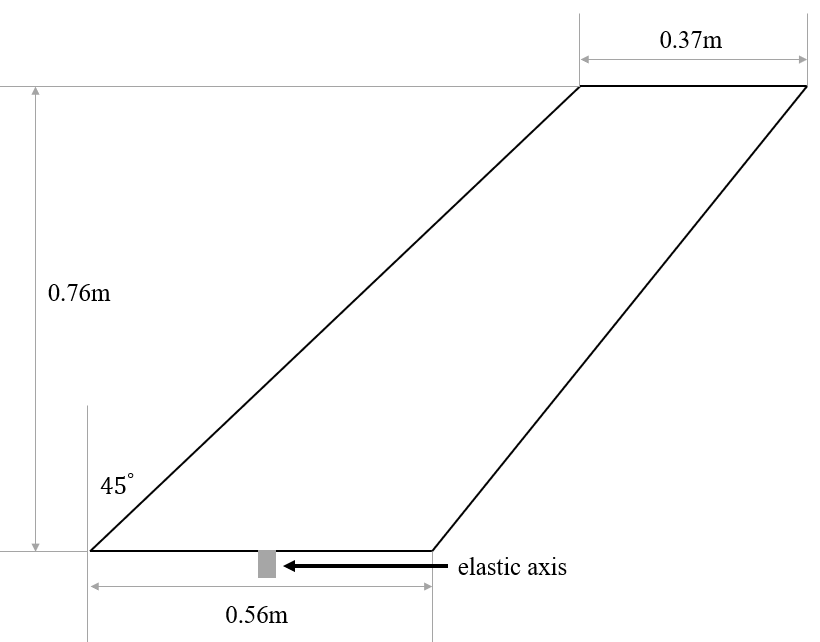}}
		\subfigure[]{\label{freeplay}
			\includegraphics[width=0.32\textwidth]{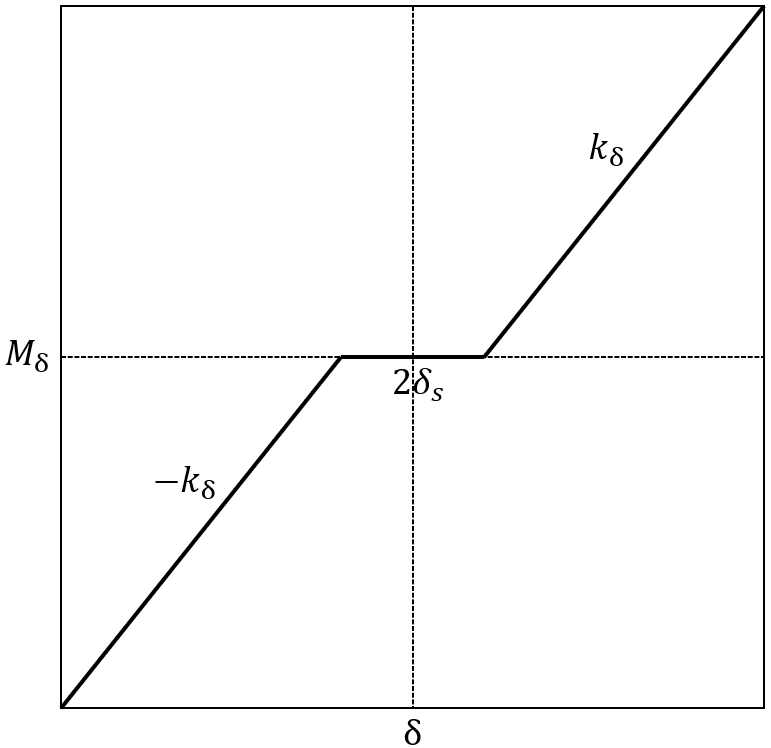}}
		\caption{a) Modified AGARD 445.6 wing geometry specifications and b) hinge stiffness as a function of rotation}
		\label{fig:1}
	\end{figure}

        The AGARD 445.6 wing is a well known transonic benchmark case, with experiments conducted in the NASA transonic dynamic wind tunnel. The model consists of a tapered swept wing (see Figure~\ref{agardBENCH}) with a NACA 65A004 airfoil section and sweep angle of 45 [$^\circ$]. The material properties considered here are those of the weakened model (No. 3)~\cite{yates85}. For comprehensive validation of the AGARD benchmark model using aerodynamic impulse responses, see recent work by the authors~\cite{hale23, candon19phd}. 
	
	In this paper, the wing is modified to represent an all-movable control surface (first presented by Carrese \textit{et al.}~\cite{carrese17}), with a torsional spring added to a node at the root, which is free to rotate about the pitch axis. The torsional spring at the root contains a zero-stiffness dead-zone and a nominal stiffness of $k_{\delta} = 500$ Nm/rad otherwise, as is depicted in Fig.~\ref{freeplay}.
	
	\subsection{Nonlinear Aeroelastic Equation-of-Motion}
	The undamped equation-of-motion for an aeroelastic system with concentrated structural nonlinearity in discrete (nodal) coordinates is given as
	\begin{equation}
		\bm{M_v}\ddot{\bm{v}} + \bm{R}(\bm{v}) + \bm{F_v} = 0 \label{eq1}
	\end{equation}
 
    \begin{equation}
      \label{eq2}
		\bm{R}({\bm{v}) = \bm{K_v}\bm{v}} + \bm{F_c}(\delta)
    \end{equation}

	\noindent where $\bm{M_v}$ and $\bm{K_v}$ are the structural mass and stiffness matrices respectively, $\bm{v} = \{v_1,\ v_2,\ \ldots,\ v_N\}^T$ is the displacement vector of $N$ degrees-of-freedom, and $\ddot{\bm{v}}$ is the second time-derivative of $\bm{v}$. $\bm{F_v} = \{F_{v1},\ F_{v2},\ \ldots,\ F_{vN}\}^T$ is the nonlinear aerodynamic force vector in nodal coordinates. The state-dependent freeplay loads are given by the term $\bm{F_c}(\delta)$ which takes the form

    \begin{equation}
    \bm{F_c}(\delta) = \left\{ \begin{array}{lllll} k_{\delta}(\delta-\delta_s) &&&& \delta > \delta_s \\ 0,  &&\text{if}&& -\delta_s < \delta < \delta_s \\ -k_{\delta}(\delta-\delta_s) &&&& \delta < -\delta_s \end{array} \right.
    \end{equation}

    \noindent where $\delta$ is the rotational displacement of the root about the freeplay hinge axis and $2\delta_s$ is the total rotational freeplay magnitude. 
 
    The system described by Eq.~\eqref{eq1} can be reduced by several orders of magnitude by considering modal coordinates, such that, the structural motion is approximated as the linear superposition of a subset of $m$ normal modes $\bm{\Phi}_{v}$ due to generalized displacement $\bm{\xi}$. Given the freeplay nonlinearity, the mode shapes in $\bm{\Phi}_{v}$ cannot properly account for localized displacements in the region of the nonlinear hinge. The fictitious masses (FM) method~\cite{karpel75} is used to improve the representation of these local deformations in the set of low frequency modes. A large fictitious mass is added to the DOF of the mass matrix where the discrepancy in localized displacements occurs, then the normal mode shapes are obtained from free vibration analysis and used in the aeroelastic simulation. The baseline FM modes $\bm{\Phi}_B$ are derived using ANSYS MAPDL, yielding the generalized system in baseline fictitious mass coordinates

    \begin{equation}
    \label{eq:modalAE}
    \bm{M}_B\bm{\ddot{\xi}} + \bm{K}_B\bm{{\xi}} +\bm{\Phi}_{B}^T\bm{F_c}(\delta)  + \bm{Q} = 0 
    \end{equation}

    \noindent where $\bm{M}_B = \bm{\Phi}_{B}^T\bm{M_v}\bm{\Phi}_{B}$ is and $\bm{K}_B = \bm{\Phi}_{B}^T\bm{K_v}\bm{\Phi}_{B}$ are the generalized mass and stiffness matrices in fictitious mass coordinates, respectively. $\bm{Q} = \bm{\Phi}_{B}^T\bm{F_v}$ is the nonlinear generalized aerodynamic force vector which is computed using the nonlinear ROM described in the previous sections or via time-marching CFD simulation (described below). The first eight eigenvalues of the baseline FM modes and baseline normal modes are given in Table~\ref{tab:fmm} where excellent agreement with the results of Carrese \textit{et al.}~\cite{carrese17} can be observed. For comprehensive numerical validation and theoretical formulation of the FM method see recent work by the authors~\cite{hale24}.

	\begin{figure}[h]
		\centering
		\subfigure[mode 1 (0Hz)]{\label{mode1_free}
			\includegraphics[width=0.23\textwidth]{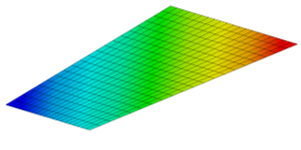}}
		\subfigure[mode 2 (28.24Hz)]{\label{mode2_free}
			\includegraphics[width=0.23\textwidth]{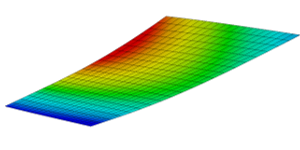}}
		\subfigure[mode 3 (39.52Hz)]{\label{mode3_free}
			\includegraphics[width=0.23\textwidth]{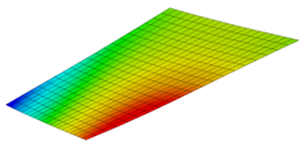}}
		\subfigure[mode 4 (84.33Hz)]{\label{mode4_free}
			\includegraphics[width=0.23\textwidth]{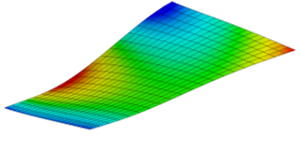}}
		\caption{First four fictitious mass modes}
		\label{fig:3}
	\end{figure}

 	\begin{table}[h]
		\centering
		\begin{tabular}{ccccc}
			\hline
                \multicolumn{2}{c}{$k_\delta = 500$ [Nm/rad]} & \multicolumn{3}{c}{$k_\delta = 0$ [Nm/rad]} \\
                \hline
			\textbf{$\omega_u$} & \textbf{$\omega_B$} & \textbf{$\omega_u$} & \textbf{$\omega_B$} & \textbf{$\omega_B$~\cite{carrese17}}  \\
			\hline
			\hline
			5.61	&	5.61	&	0.000	&	0.001 & 0.0001\\
			31.93	&	31.93	&	28.24	&	28.24 &	28.24\\
			39.52	&	39.52	&	39.52	&	39.52	& 39.52\\
			88.03	&	88.04	&	84.33	&	84.35 &	84.36\\
			96.56	&	96.56	&	96.53	&	96.53	& 96.53\\
			131.71	&	131.71	&	130.84	&	130.93 &	131.01\\
			133.90	&	133.97	&	131.71	&	131.71	& 131.79\\
			167.21	&	814.91	&	167.21	&	745.24 &	700.79\\
			\hline
		\end{tabular}
		\caption{Natural frequencies for the root stiff and free cases calculated directly with fictitious masses}
		\label{tab:fmm}
	\end{table}
	
	At this point it is important to note that in the general definition of the Taylor series expansion of the unsteady aerodynamic forces, the structural displacements defined by $\bm{u}$ (Eq.~\ref{eq:TS2}) are equivalent to $\bm{\xi}$ and will be referred to as such from now on.

	\subsection{Computational Fluid Dynamics Model}
	For the FOM, the generalized aerodynamic force vector $\bm{Q}$ is obtained using the commercial finite-volume Navier-Stokes solver ANSYS Fluent 2023 R1. The Euler equations for transient flowfields are solved via a coupled pressure-based solver with implicit second-order spatial and first-order temporal discretization of the flowfields with Rhie-Chow: distance-based flux interpolation. The convergence criteria are set to $1\times10^{-4}$ for the scaled residuals at each time-step. The investigation is conducted on a structured grid of $70\times10^{3}$ elements, with a minimum orthogonal quality of 0.032. It is important to note that this numerical mesh is validated against experimental campaign~\cite{yates85} via linear stability analysis~\cite{candon19phd, hale23} for the unmodified AGARD wing. Grid deformation is facilitated using a diffusion-based approach. The Modal Projection and force Reconstruction (MPR) method~\cite{joseph21} is used to project the structural mode shapes onto the fluid grid. MPR includes a robust interpolation scheme that accounts for disparity in the grid topologies, and conserves forces and moments. 
	

    \subsection{Nonlinear Aerodynamic Reduced Order Model}

\begin{figure}[h]
        \centering
        \includegraphics[width=1\textwidth]{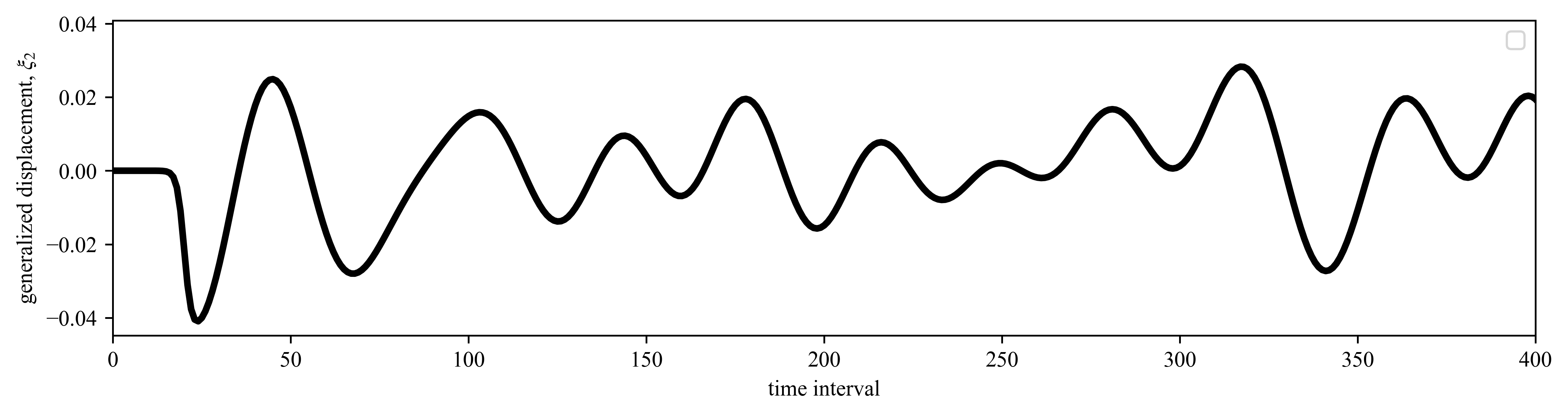}
        \caption{Band limited random excitation used for mode 1}
        \label{fig:aeroInp}
    \end{figure}

    In this paper the ROMs are generate with approximate knowledge of the modal LCO response for $\delta_s = \pm 1^\circ$, using the various approaches described in Section~\ref{sec:roms}. Table~\ref{tab:aerorandom} summarizes the frequencies and maximum amplitudes of the random excitation functions for each mode. Four modes are used in in the identification procedure. Fig.~\ref{fig:aeroInp} presents an example of the band limited random excitation. A total of $n = 400$ samples are generated for each mode. 
        
    \begin{table}[h]
    \centering
    \begin{tabular}{ccccccccc}
        \hline
          \multicolumn{2}{c}{Mode 1} & \multicolumn{2}{c}{Mode 2} & \multicolumn{2}{c}{Mode 3}& \multicolumn{2}{c}{Mode 4}  \\
        \hline
           $f$ [Hz] & $\xi_{1,max}$ & $f$ [Hz] & $\xi_{2,max}$& $f$ [Hz] & $\xi_{3,max}$& $f$ [Hz] & $\xi_{4,max}$  \\
        \hline
            \hline
             0-25 & 0.04 & 0-30 & 0.03 & 0-30 & 0.03 & 0-100 & 0.01 \\
        \hline
    \end{tabular}
    \caption{Parameters of the band limited random excitation}
    \label{tab:aerorandom}
    \end{table}

     
        
        


     \subsection{Aeroelastic Time Integration}
     
    The aeroelastic system is solved using the RMIT in-house Fluid-Structure Interaction code PyFSI. Aeroelastic solutions are achieved by marching Eq.~\ref{eq:modalAE} forward in time, where the wing transient structural motion is solved using Newmark-$\beta$ time-integration. Newton-Raphson iterations are used to converge the state-dependent freeplay load within each time-step by minimizing error in the stiffness matrix. For the FOM, $\bm{Q}$ in Eq.~\ref{eq:modalAE} is resolved using the CFD model described above, solving for the nonlinear fluid loads at every time-step. For the aerodynamic ROM, the same term is resolved by solving the forward linear problem described in Eq.~\ref{eq:linprob} at each time-step. A time-step of $\Delta t = 0.001$ s is used. 
        
    \section{Results and Discussion}
    \label{sec:res}
    
    In this section the aeroelastic responses of the various ROMs for the different freeplay values, velocity indexes, and Mach numbers are presented and discussed. In Sections A-F, the freestream Mach number is $M_\infty = 0.96$, AoA is $\alpha_0 = 0^\circ$ and the velocity index $V^*$ is set to 96\% of the linear flutter speed, $V^* = 0.96V^*_f = 0.192$. In Section G, the velocity index is varied, and ROMs are generated for $M_\infty = 0.901$ and $M_\infty = 0.96$.

    \subsection{Performance Metrics}

    Some performance metrics are now explicitly defined, given that this section considers two separate optimization problems. Firstly, the normalized root mean square deviation (nrmsd) is used to quantify the error between the FOM and ROM, defined for a vector of length $n$ according to

    \begin{equation}
        nrmsd(FOM,ROM) = \frac{100\sqrt{\sum_{i=0}^n (FOM(i)-ROM(i))^2}}{FOM_{max}-FOM_{min}}
    \end{equation}
    
    The aerodynamic optimization problem uses the objective $\mathrm{argmin}||\bm{Q_{FOM}}(\xi) -\bm{Q_{ROM}}(\xi)||$ which is quantified using a weighted function. The aerodynamic response in each of the four modes is given a weight according to an approximation of the relative amplitude of the mode in the LCO, according to 

    \begin{equation}
    \begin{split}
        nrmsd_Q = \sum_{j=1}^{4} \left(
       0.8nrmsd(\bm{Q_{FOM}^{1j}},\bm{Q_{ROM}^{1j}}) + 0.07 nrmsd(\bm{Q_{FOM}^{2j}},\bm{Q_{ROM}^{2j}}) \right. \\ \left.
        + 0.11nrmsd(\bm{Q_{FOM}^{3j}},\bm{Q_{ROM}^{3j}})
        + 0.02nrmsd(\bm{Q_{FOM}^{4j}},\bm{Q_{ROM}^{4j}}) \right) /4
    \end{split}
    \end{equation}

    \noindent where, for example, $\bm{Q_{FOM}^{1j}}$ is the FOM aerodynamic response for $\bm{Q_1}(\bm{\xi_j})$. The aeroelastic optimization problem uses the objective $\textrm{argmin}||\delta_{FOM}(t) - \delta_{ROM}(t)||$ which is quantified according to

    \begin{equation}
     nrmsd_{\delta} = nrmsd(\delta_{FOM}(t),\delta_{ROM}(t)) 
    \end{equation}

    \noindent where $\delta$ is the rotational aeroelastic response at the root hinge node. The performance is also assessed using a sparsity penalization term, defined according to

        \begin{equation}
        \text{$s$-score} = nrmsd \times \frac{n_{s}}{\sqrt{n_\kappa}}
        \end{equation}

        \noindent where $n_s$ is the number of non-zero coefficients in the ROM (all $m \times m$ modes) compared to $s$ which is the number of non-zero coefficients in single partial derivative tensor. Similarly $n_\kappa$ is the total number of coefficients in the ROM (all $m \times m$ modes). The sparsity score penalizes the nrmsd according to the ratio of the number of non-zero coefficients to the square root of the total number of coefficients. 

    \subsection{Nonlinear Aerodynamic ROM Performance}
    
    To demonstrate the benefits of sparsity promotion, convergence studies are conducted for the generalized aerodynamic forces, comparing $nrmsd_Q$ for the linear, full third-order and sparse third-order ROMs. The performance is assessed on a separate cross-validation dataset of 400 samples. 
    
\begin{figure}[h]
            \centering
            \includegraphics[width=1\textwidth]{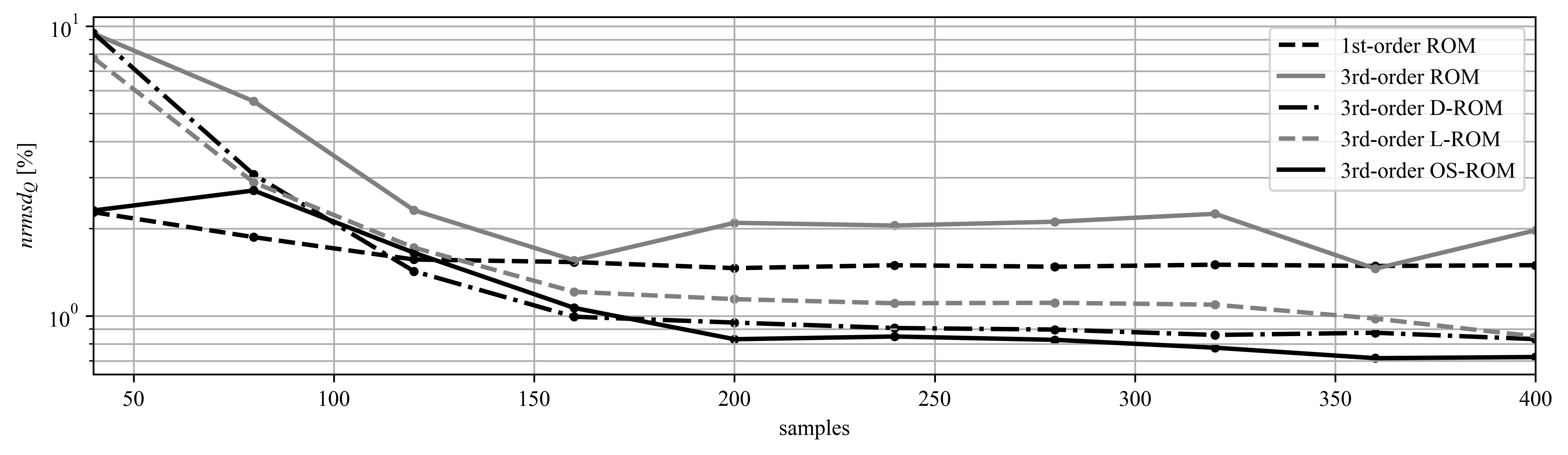}
            \caption{Convergence of the aerodynamic ROMs}
            \label{fig:aeroConv}
        \end{figure}

    Figure~\ref{fig:aeroConv} demonstrates that the linear model appears to converge with $n \approx 100-150$ samples and the third-order OS-ROM and D-ROM with $n \approx 150-200$ samples. The third-order L-ROM requires more samples to converge, which is expected given that the number of non-zero coefficients is larger. On the other hand, the full third-order ROM is yet to converge with $n = 400$ samples at which point the performance remains worse than the linear model. The OS-ROM demonstrates superior performance to all other sparsity promoting techniques.

\begin{figure}[h]
            \centering
            \includegraphics[width=1\textwidth]{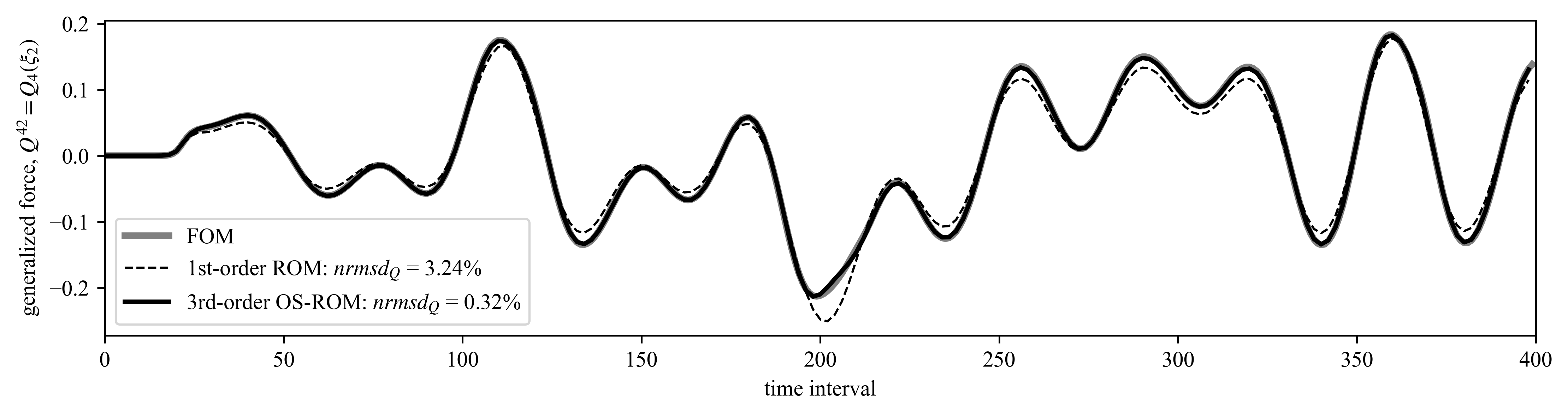}
            \caption{Comparison between the FOM and OS-ROM generalized aerodynamic forces for $\bm{Q_4}(\bm{\xi_2})$}
            \label{fig:aeroSeries}
        \end{figure}

    Figure~\ref{fig:aeroSeries} presents an example of the generalized aerodynamic forces in mode 4 due to perturbation of mode 2, comparing $\bm{Q_{FOM}^{42}}$, the linear $\bm{Q_{ROM}^{42}}$ and the sparse third-order $\bm{Q_{OS-ROM}^{42}}$. It is clear that the third-order OS-ROM provides superior performance with a reduction in error from $nrmsd_Q = 3.42\%$ to  $nrmsd_Q = 0.32\%$, nearly perfectly overlaying the FOM aerodynamic response. Detailed aerodynamic and aeroelastic hyperparameter tuning is conducted and will be discussed in the following section and in the Appendix.
        
    \subsection{Hyperparameter Tuning}

      The hyperparameter tuning is initially conducted to minimize the error between the FOM and ROM aerodynamic response $nrmsd_Q$, $i.e.$, using only the aerodynamic cross-validation dataset. In this way, the application of the aerodynamic optimized ROM to the aeroelastic problem is a test of out-of-sample performance. Hyperparameter tuning is also conducted for the aeroelastic response using $nrmsd_\delta$, $i.e.$, tuning the ROM for optimal aeroelastic performance with $\delta_s = \pm 1^{\circ}$. The grid search parameters for the various ROMs are summarized in Table~\ref{tab:HP}.  The results of the grid search are presented in the Appendix.

\begin{table}[h]
    \centering
        \begin{tabular}{ccccl}
        \hline
        \textbf{type} & $k$ & $s$ &  $\alpha$ & Notes\\
        \hline
        \hline
        ROM & 5-20 & - & - & No sparsity\\
        D-ROM & 5-20 & - & - & Rigid sparsity $n_s = pk$ \\
        L-ROM & 5-20  & - & $5\times 10^{-14}$ - $1\times 10^{-9}$ & Sparsity via LASSO  \\
        OS-ROM & 5-20  & 5-25 &- & Sparsity via OMP  \\
	\hline
	\end{tabular}
    \caption{Grid search hyperparameters for each ROM}
    \label{tab:HP}
\end{table}

\subsection{Aeroelastic Response using the Aerodynamic-Optimized ROM}

The out-of-sample performance of the ROMs is now assessed by applying the aerodynamic-optimized ROM to the aeroelastic problem. The naming convention for the ROMs describes the hyperparameter optimization problem (aerodynamic or aeroelastic) and the ROM order, $e.g.$, AD1 denotes a first-order aerodynamic-optimized ROM and AE3 denotes a third-order aeroelastic-optimized ROM. The linear ROM(AD1) and ROM(AE1) underwent hyperparameter tuning consistent with that described in the previous section, as did the full third-order ROM(AD3) and ROM(AE3).

\begin{figure}[h]
		\centering
                \subfigure[]{\label{mode1_free}
			\includegraphics[width=0.32\textwidth]{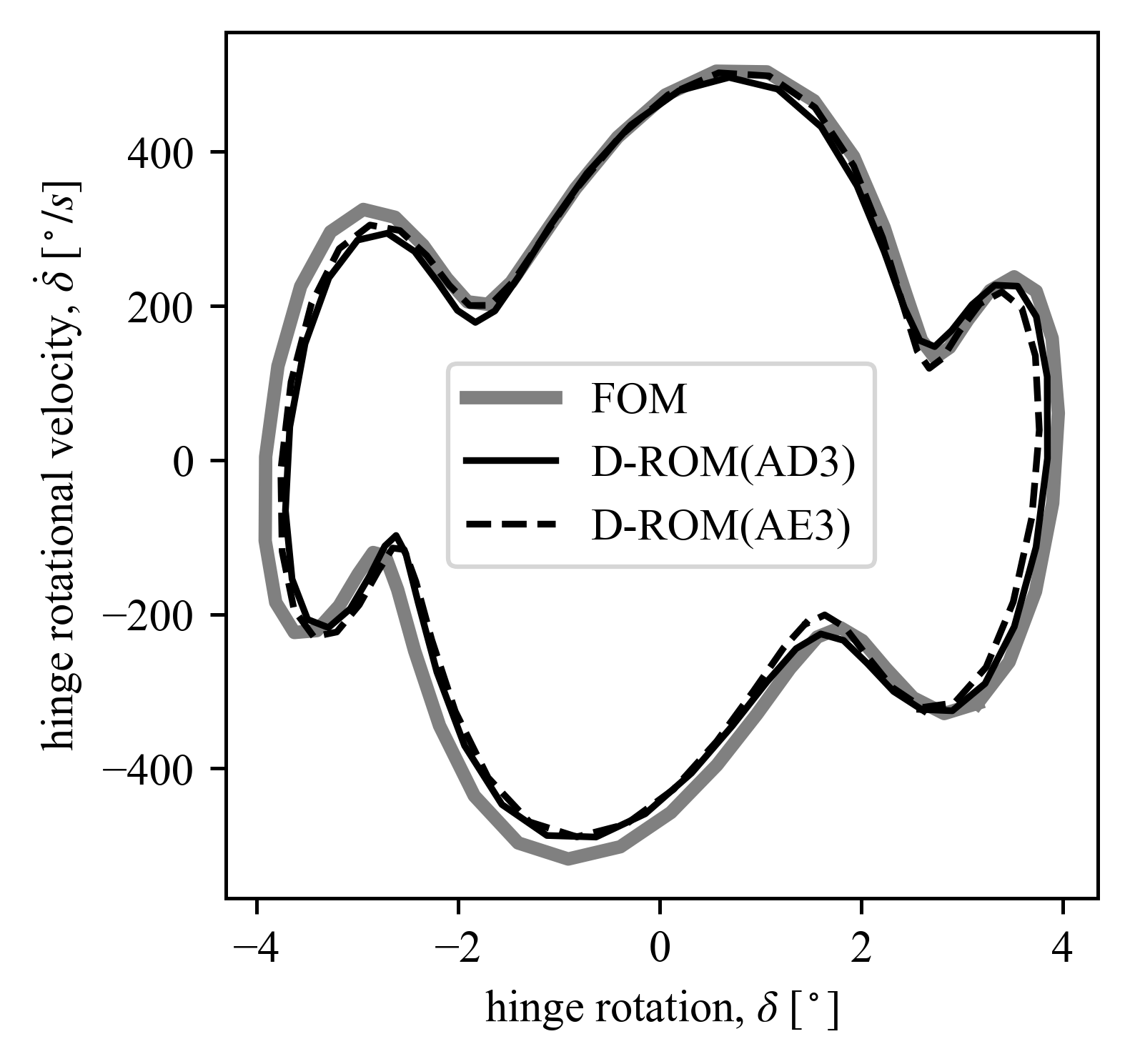}}
                \subfigure[]{\label{mode1_free}
			\includegraphics[width=0.32\textwidth]{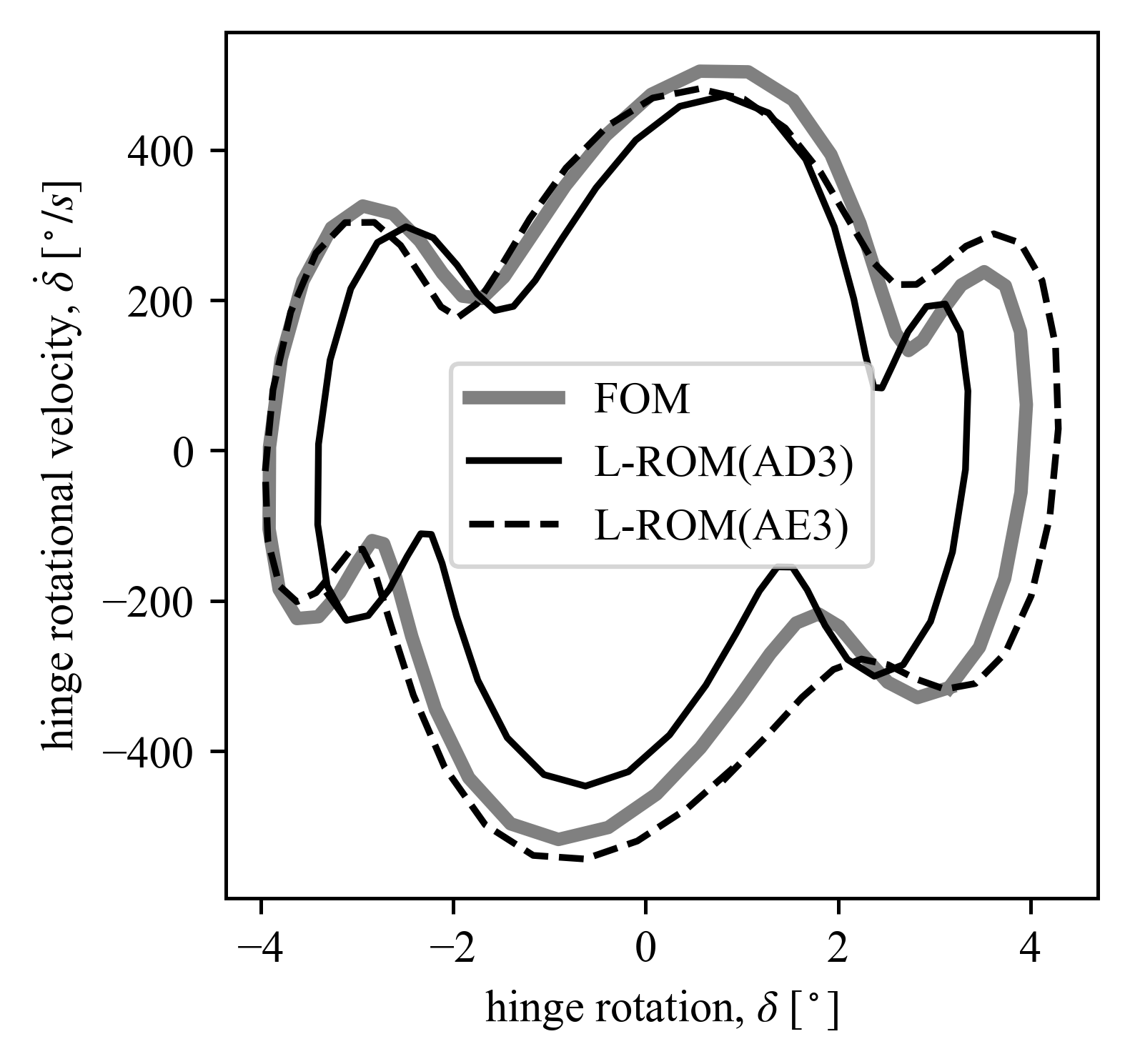}}
                \subfigure[]{\label{mode1_free}
			\includegraphics[width=0.32\textwidth]{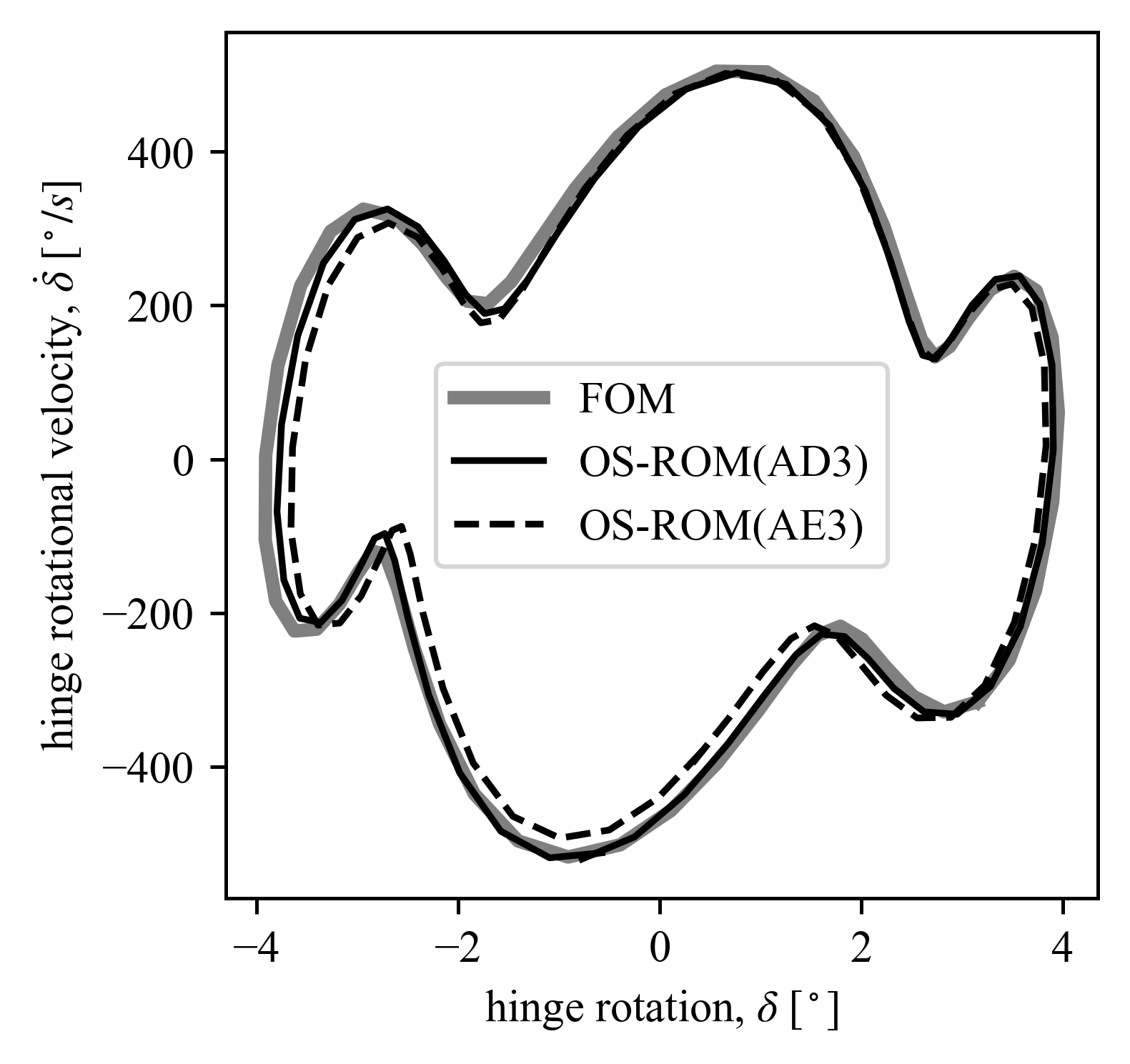}}

                \subfigure[]{\label{mode1_free}
			\includegraphics[width=0.32\textwidth]{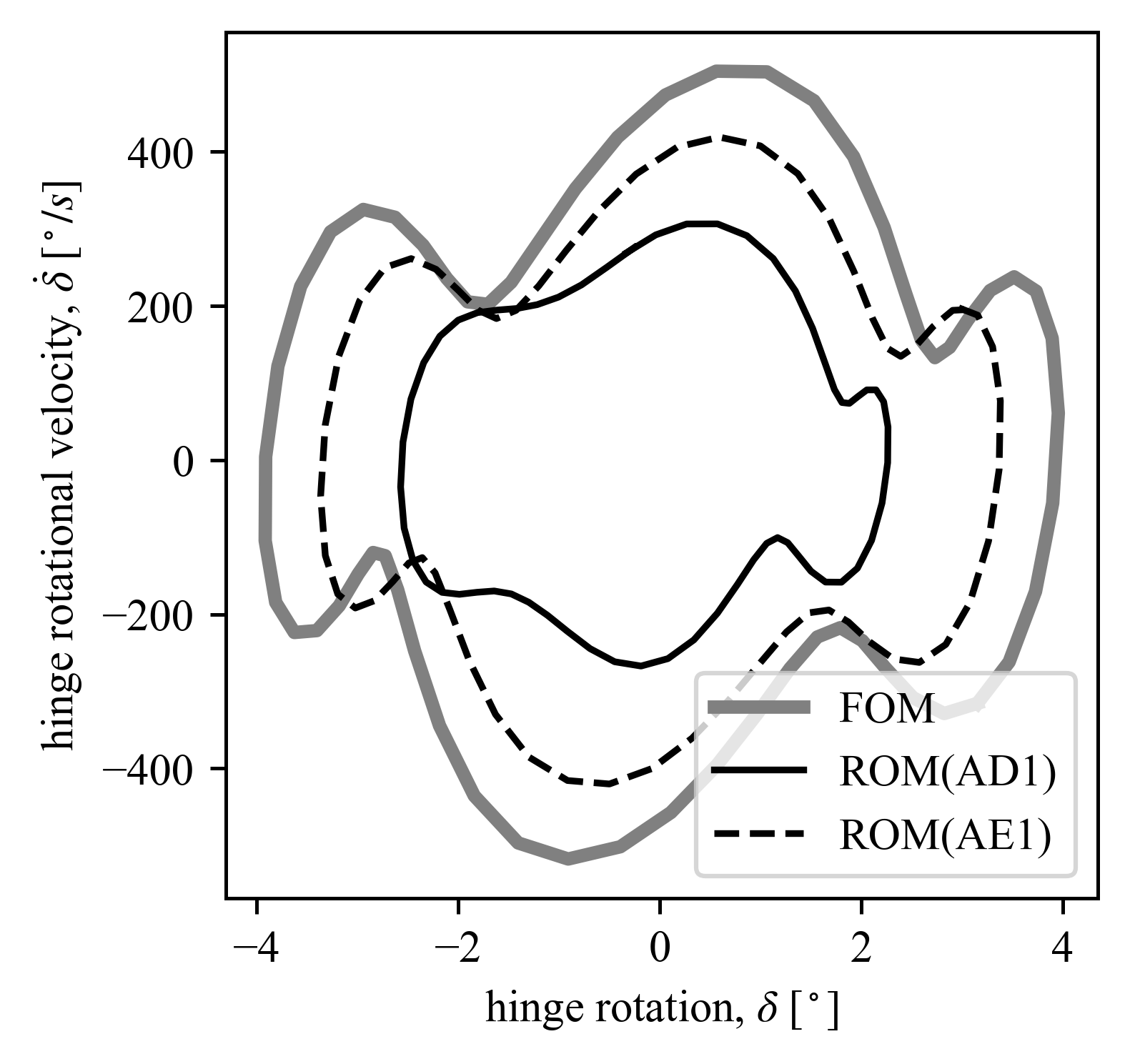}}
                \subfigure[]{\label{mode1_free}
			\includegraphics[width=0.32\textwidth]{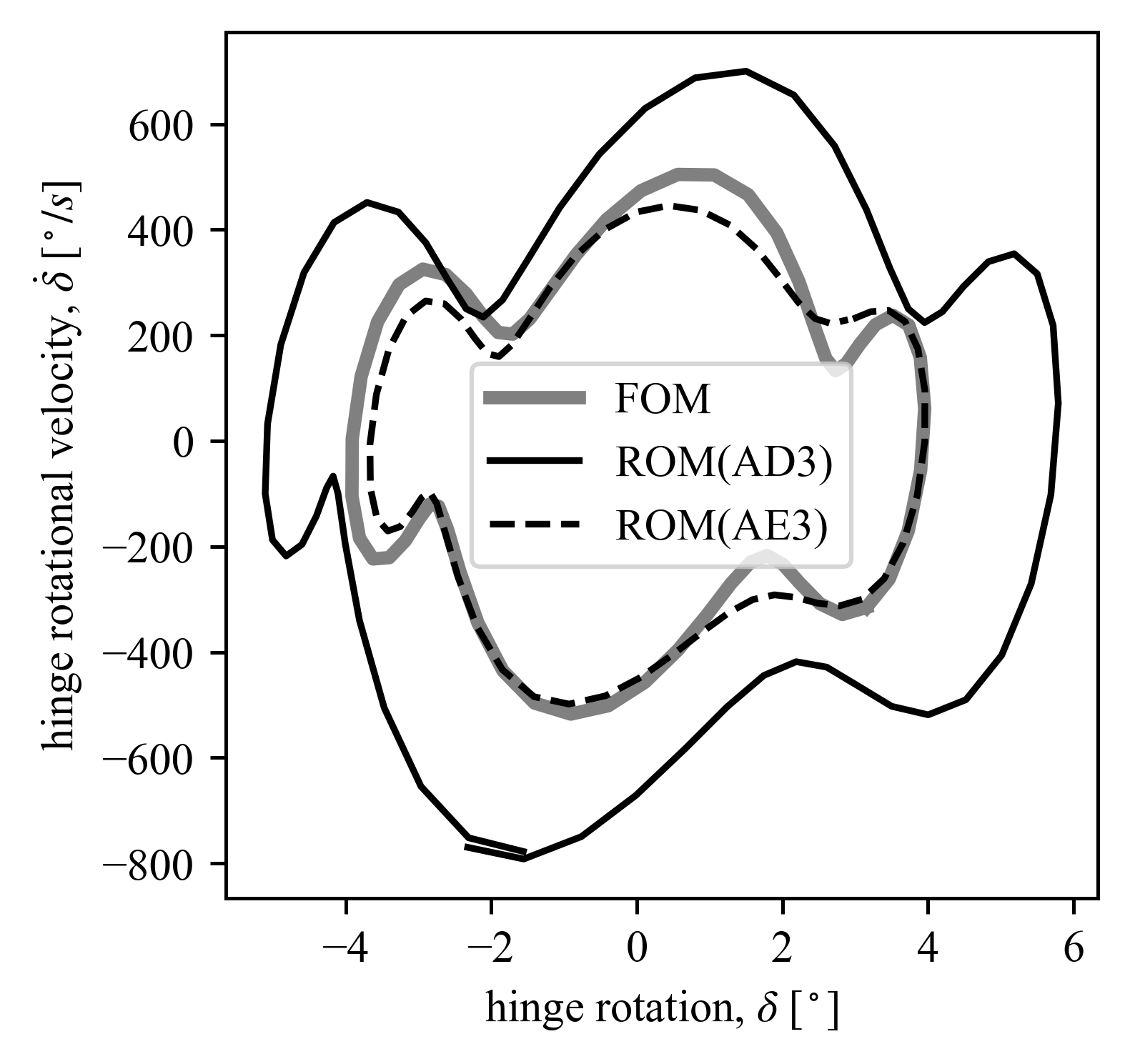}}
		\caption{Phase portraits comparing the FOM and ROM solutions with $\delta_s = \pm  1^\circ$ }
		\label{fig:opt_res}
	\end{figure}

Table~\ref{tab:opt_res} and Fig.~\ref{fig:opt_res} present the results of the hyperparameter optimization. Firstly, considering the ROMs without sparsity, it can be seen that the linear ROM(AD1) and full third-order ROM(AD3) are both unable predict the limit cycle accurately, with errors of 13.81\% and 15.46\%, respectively. Furthermore, the $s$-score is high for both given that neither consider sparsity. Observation of the corresponding aeroelastic-optimized ROM(AE1) and ROM(AE3) indicates that the performance can be improved significantly by tuning the ROM for the aeroelastic case, however, still neither are able to capture the LCO with excellent precision, $i.e.$, significant under prediction is observed for ROM(AE1) and the general form cannot be captured by ROM(AE3). These findings demonstrate that linearization of the nonlinear aerodynamic forces yields poor performance and that the length of the training data is insufficient to identify the full third-order tensors of Taylor partial derivatives - highlighting the need for sparsity promotion.

Of the ROMs generated with sparsity, excellent out-of-sample performance can be observed for the aerodynamic-optimized D-ROM(AD3) and OS-ROM(AD3) with both achieving the target of $<2\%$ error. Only marginal improvements can be achieved by tuning the D-ROM(AE3) and OS-ROM(AE-3) for optimal aeroelastic performance. Although both perform well in terms of the nrmsd metric, the $s$-score must be taken into account, demonstrating that OS-ROM(AD3) and OS-ROM(AE3) outperform D-ROM(AD3) and D-ROM(AE3) given the significantly lower number of non-zero coefficients in the ROMs. Furthermore, the close proximity of the global minima and general similarity in the heatmaps for OS-ROM(AD3) and OS-ROM(AE3) (see Appendix) provides strong evidence that the aerodynamic-optimized OS-ROM can generalize to the aeroelastic case. While, on the other hand, given the significant distance in the hyperparameter space between the global minima for the D-ROM(AD3) and D-ROM(AE3) (see Appendix), the excellent performance of D-ROM(AD3) may be anomalous and does not necessarily confirm the ability to generalize. 

The aerodynamic-optimized L-ROM(AD3) performance is poor, suggesting that $\ell_1$ regularization is not appropriate for this framework, $i.e.$, given the small training budget, it is preferable to have control over the number of non-zero coefficients using a greedy optimization algorithm such as OMP, than it is to use an $\ell_1$ penalization term in least squares optimization. This is further supported by the large $s$-scores that occur for both L-ROM(AD3) and L-ROM(AE3), given that the number of non-zero coefficients large.

\begin{table}[h]
    \centering
    \begin{tabular}{ccccccccccc}
    \hline
         name & alg. & obj. & $p$ & $k$ & $s$ &  $\alpha$  & $nrmsd_\delta$ [\%] & $n_{s}$ & $n_\kappa$ & $s$-score\\
        \hline
        \hline
            ROM(AD1)& LS & $nrmsd_Q$ & 1 & 11 & - & - & 13.81 & 176 & 176& 31.9 \\
            ROM(AE1)&LS &$nrmsd_\delta$ & 1 & 20 & - & - & 3.88 & 320 & 320& 7.4\\
            ROM(AD3)&LS &$nrmsd_Q$ & 3 & 11 & - & - & 15.46 & 5808& 5808 & 1178.2 \\
            ROM(AE3)&LS &$nrmsd_\delta$  & 3 & 13 & - & - & 2.93 & 8944 & 8944 & 277.1 \\
            D-ROM(AD3)& LS &$nrmsd_Q$ &3 & 25 & - & - & 1.72 & 1200 & 52400 & 9 \\
            D-ROM(AE3)& LS & $nrmsd_\delta$  &3& 16 & - & - & 0.91 & 768 & 15488 & 5.6 \\
            L-ROM(AD3)& LASSO &$nrmsd_Q$ & 3 & 9 & - & $1\times 10^{-11}$ & 7.56 & 1847 & 3504 & 235.8 \\
            L-ROM(AE3)& LASSO & $nrmsd_\delta$  &3& 11 & - & $5\times 10^{-10}$ & 1.37 & 716 & 5808 & 12.9 \\
            OS-ROM(AD3)&OMP & $nrmsd_Q$ &3& 12 & 14 & - & 1.28 & 224 & 7264 & 3.4\\
            OS-ROM(AE3)&OMP & $nrmsd_\delta$  &3& 12 & 17 & - & 1.06 & 238 & 7264& 2.9\\
	\hline
    \end{tabular}
\caption{Hyperparameter optimization results for the various ROMs with aerodynamic and aeroelastic objectives}
\label{tab:opt_res}
\end{table} 
\clearpage

    \subsection{Transonic Aerodynamic Nonlinearity}

    Before investigating the ability of the aeroelastic-optimized ROMs to generalize, the transient flowfields are investigated to assess the presence and strength of transonic aerodynamic nonlinearity at the minimum and maximum freeplay values of interest.
    
\begin{figure}[h]
		\centering
            \subfigure[$0T$]{\label{mode1_free}
			\includegraphics[width=0.32\textwidth]{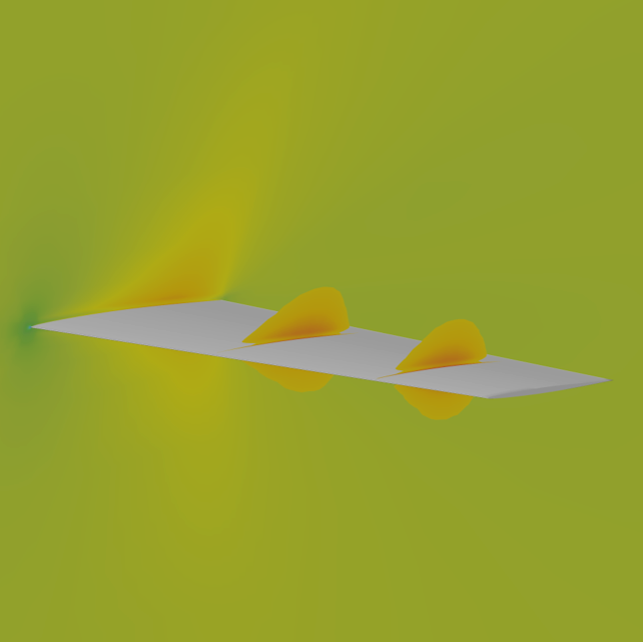}}
		\subfigure[$0.25T$]{\label{mode2_free}
			\includegraphics[width=0.32\textwidth]{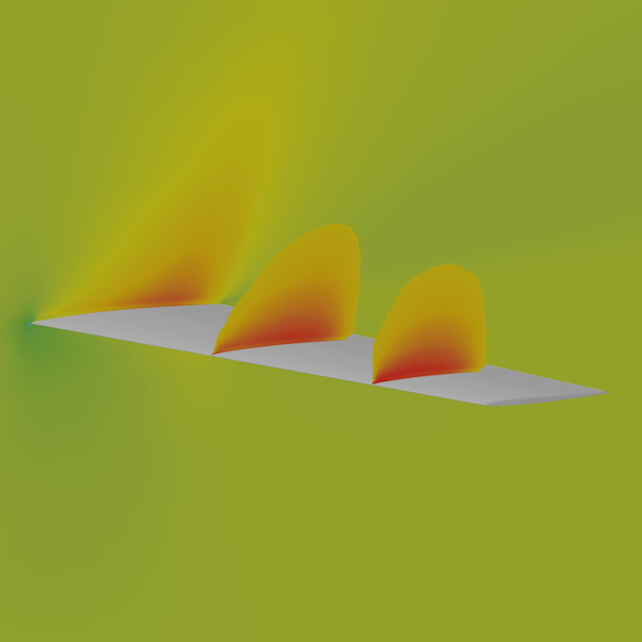}}
            \subfigure[$0.75T$]{\label{mode2_free}
			\includegraphics[width=0.32\textwidth]{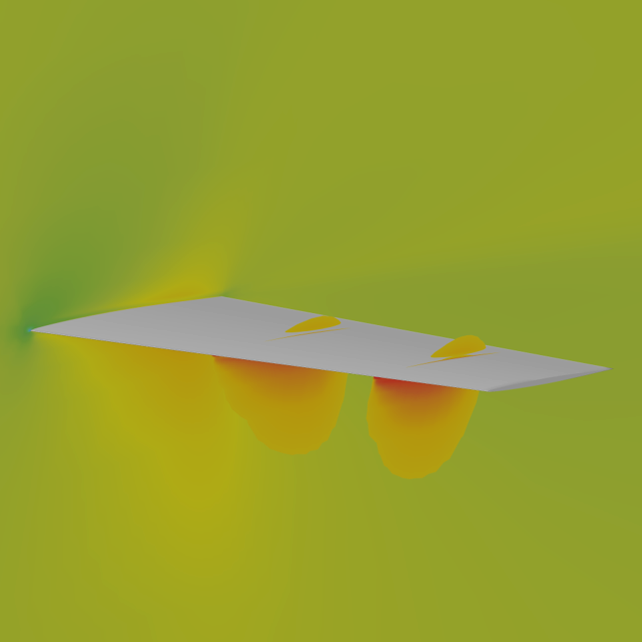}}
            \includegraphics[width=0.7\textwidth]{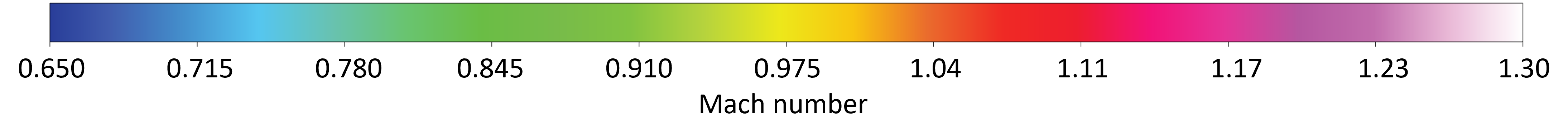}
		\caption{Mach number flowfields at the a) mid-point, b) peak, and c) trough of the limit cycle with $\delta_s = \pm 0.5^\circ$}
		\label{fig:mach05}
	\end{figure}

\begin{figure}[h]
		\centering
   
		\subfigure[$0T$]{\label{mode1_free}
			\includegraphics[width=0.32\textwidth]{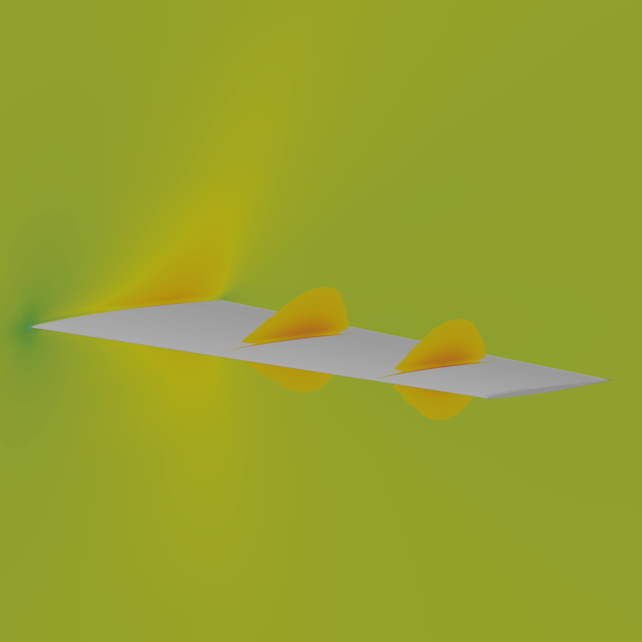}}
		\subfigure[$0.25T$]{\label{mode2_free}
			\includegraphics[width=0.32\textwidth]{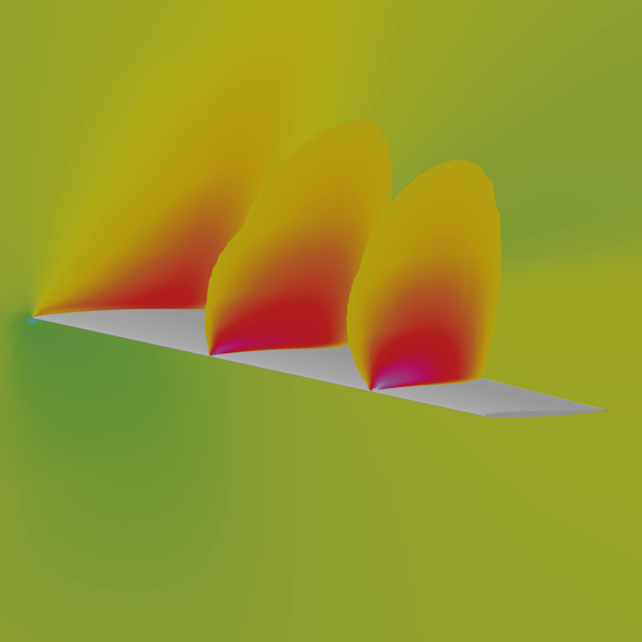}}
            \subfigure[$0.75T$]{\label{mode2_free}
			\includegraphics[width=0.32\textwidth]{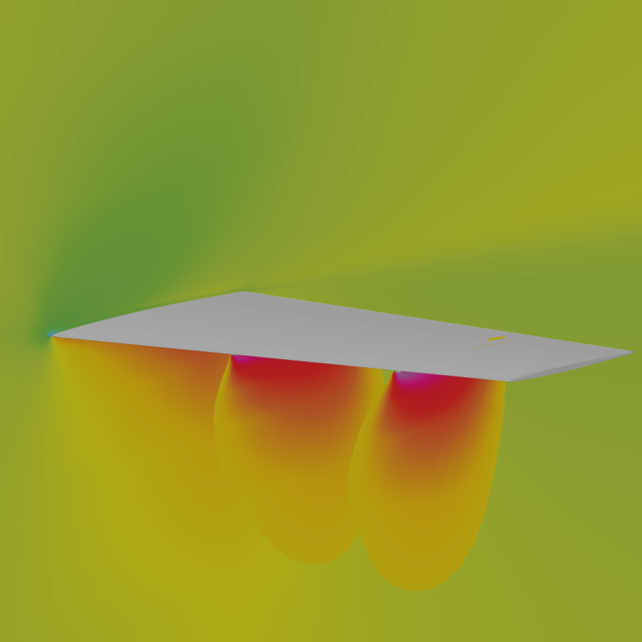}}

            \includegraphics[width=0.7\textwidth]{Figures/colorbar.PNG}
		\caption{Mach number flowfields at the a) mid-point, b) peak, and c) trough of the limit cycle with $\delta_s = \pm 1.25^\circ$}
		\label{fig:mach125}
	\end{figure}

    Figures~\ref{fig:mach05} and~\ref{fig:mach125} present the Mach number flowfields at the mid-point, peak and trough of the LCO with $\delta_s = \pm 0.5^\circ$ and $\delta_s = \pm 1.25^\circ$, respectively. For $\delta_s = \pm 0.5^\circ$ the maximum rotation about the hinge axis is $\delta \approx \pm 1.9^\circ$. Mild aerodynamic nonlinearity can be observed in the form of inviscid shock wave structures and motion. The shock waves do not completely disappear on the surface opposing the direction of the rotation. For $\delta_s = \pm 1.25^\circ$ the rotation about the hinge axis is $\delta \approx \pm 4.2^\circ$ and clearly characterized by significantly stronger aerodynamic nonlinearity. Specifically, the transonic shock wave dynamics at the extrema exhibit a stronger shock forming (which gets weaker from tip to root) and complete disappearance of the shock wave on the opposing surface. 

    This significant variation in the nonlinear aerodynamic behavior, ranging from mild transonic shock wave behavior at $\delta_s = \pm 0.5^\circ$ to more severe variance in the shock wave structure and dynamics at $\delta_s = \pm 1.25^\circ$, suggests that it is a reasonable test of the ability of a single aeroelastic optimized ROM to generalize.

    \clearpage

    \subsection{Aeroelastic Response for Varying Freeplay Values}
    The aeroelastic-optimized ROM out-of-sample performance can be evaluated by varying the freeplay value, $i.e.$, tuning the ROM for optimal aeroelastic performance is a valid strategy provided that the ROM will generalize to new points in the parameter space. The full third-order ROM(AE3) is not considered here and the first-order ROM(AE1) is used as a baseline.

\begin{figure}[h]
		\centering
			\includegraphics[width=0.66\textwidth]{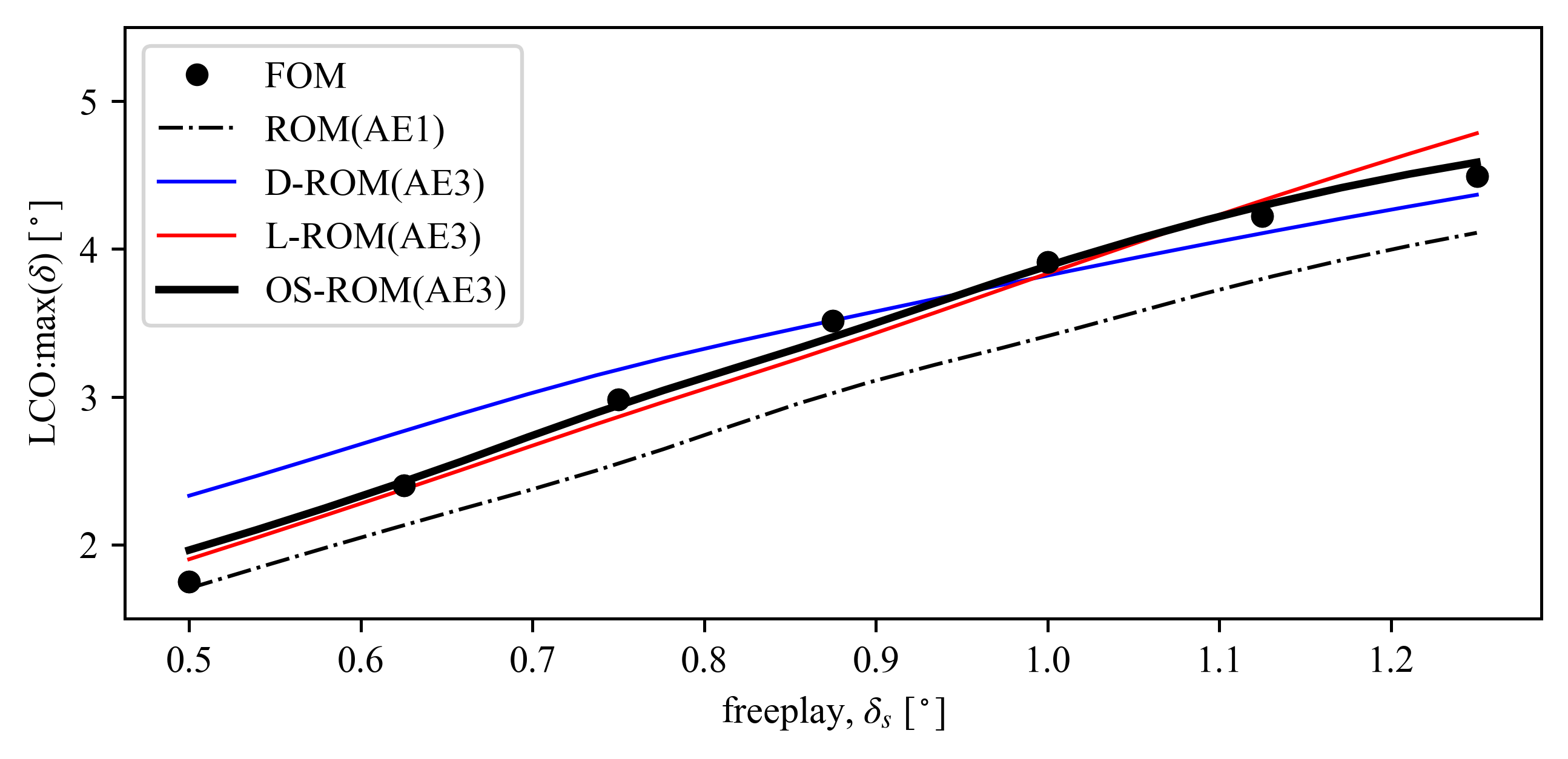}
		\caption{LCO amplitude for different freeplay values comparing the FOM to the various aeroelastic-optimized ROM solutions}
		\label{fig:fp_sweep}
	\end{figure}

    Figure~\ref{fig:fp_sweep} presents the LCO amplitude for freeplay values ranging from $\delta_s = \pm 0.5^\circ$ - $\delta_s = \pm 1.25^\circ$ with each ROM tuned to $\delta_s = \pm 1^\circ$. The linear ROM(AE1) performs well for $\delta_s = \pm 0.5^\circ$, however, consistently under predicts the LCO amplitude for larger freeplay values. The other ROM which performs poorly is the D-ROM(AE3) which, recalling that it performed very well for both optimization problems, is surprising. This is a clear example of over fitting and demonstrates the need for optimal sparsity identification.

    The L-ROM(AE3) performs reasonably for $\delta_s = \pm 0.625^\circ - 1.125^\circ$, however, it is essentially demonstrating a linear increase in amplitude and is unable to capture the nonlinear aerodynamic damping effects that occur at higher freeplay values. It was also shown in the previous section that this ROM presents some inaccuracies in the form of the LCO.

    The OS-ROM(AE3) performance is generally good, capturing the nonlinear aerodynamic damping for larger freeplay values and performing well down to $\delta_s = \pm 0.625^\circ$. Over prediction of the amplitude can be observed for $\delta_s = \pm  0.5^\circ$.  

\begin{figure}[h]
		\centering
			\includegraphics[width=0.7\textwidth]{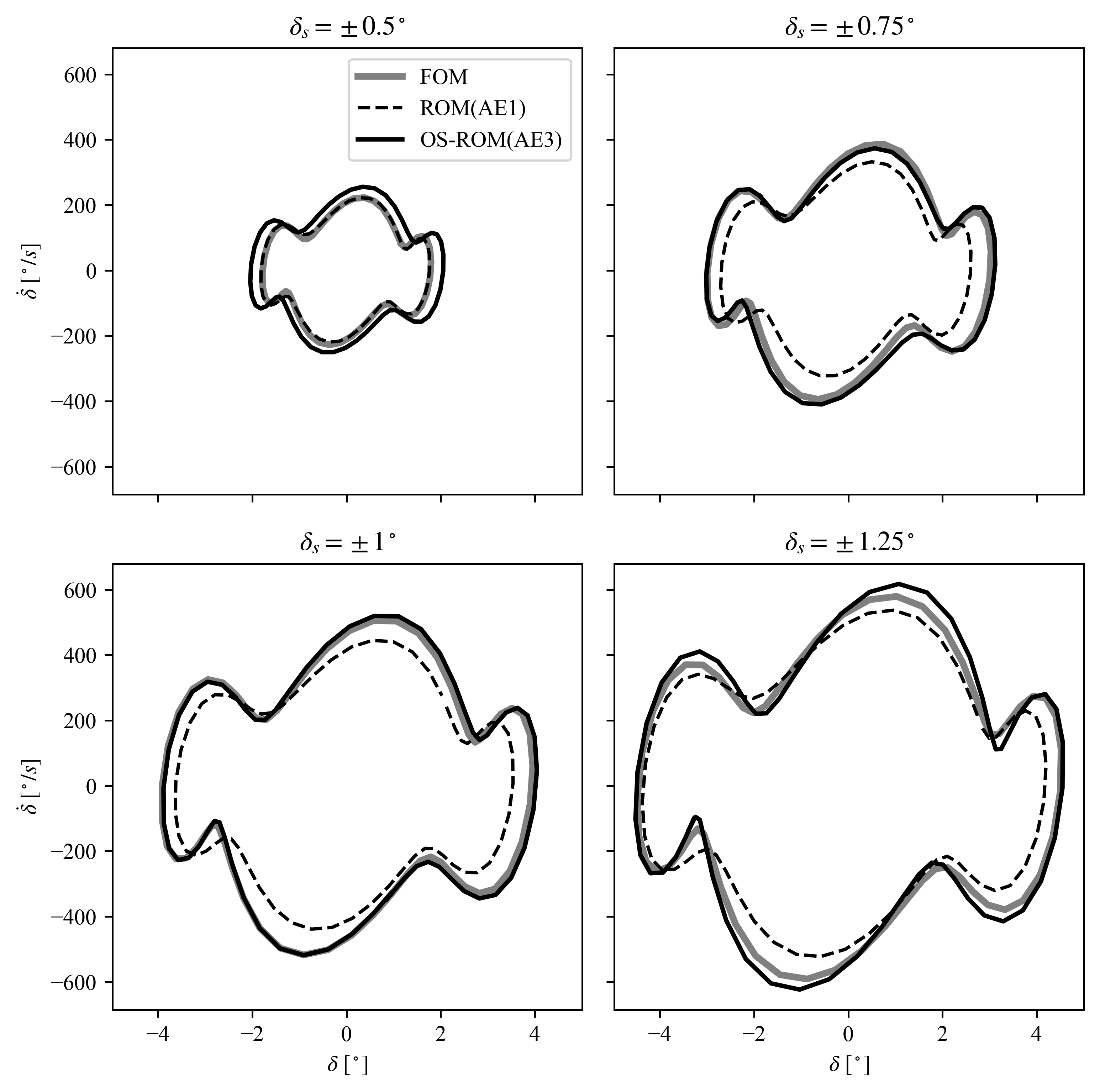}
		\caption{Phase portraits for different freeplay values comparing the FOM, ROM(AE1) and OS-ROM(AE3) solutions}
		\label{fig:phase_aeopt}
	\end{figure}

    The phase portraits for the linear ROM(AE1) and OS-ROM(AE3) are presented in Fig.~\ref{fig:phase_aeopt}. It is shown that ROM(AE1) is able to capture the nonlinear form of the LCO well for all freeplay values (although consistently under predicting the amplitude). Indeed, for some applications, this could be considered sufficiently accurate. It is important to make the distinction at this point that the linear ROM(AE1) is not equivalent to traditional linearization using impulses~\cite{silva97}. The OS-ROM(AE3) performance is very good for freeplay values greater than $\delta_s = \pm 0.5^\circ$ as is the ability to capture the nonlinear form of the LCO for all freeplay values.


\begin{figure}[h]
		\centering
			\includegraphics[width=1\textwidth]{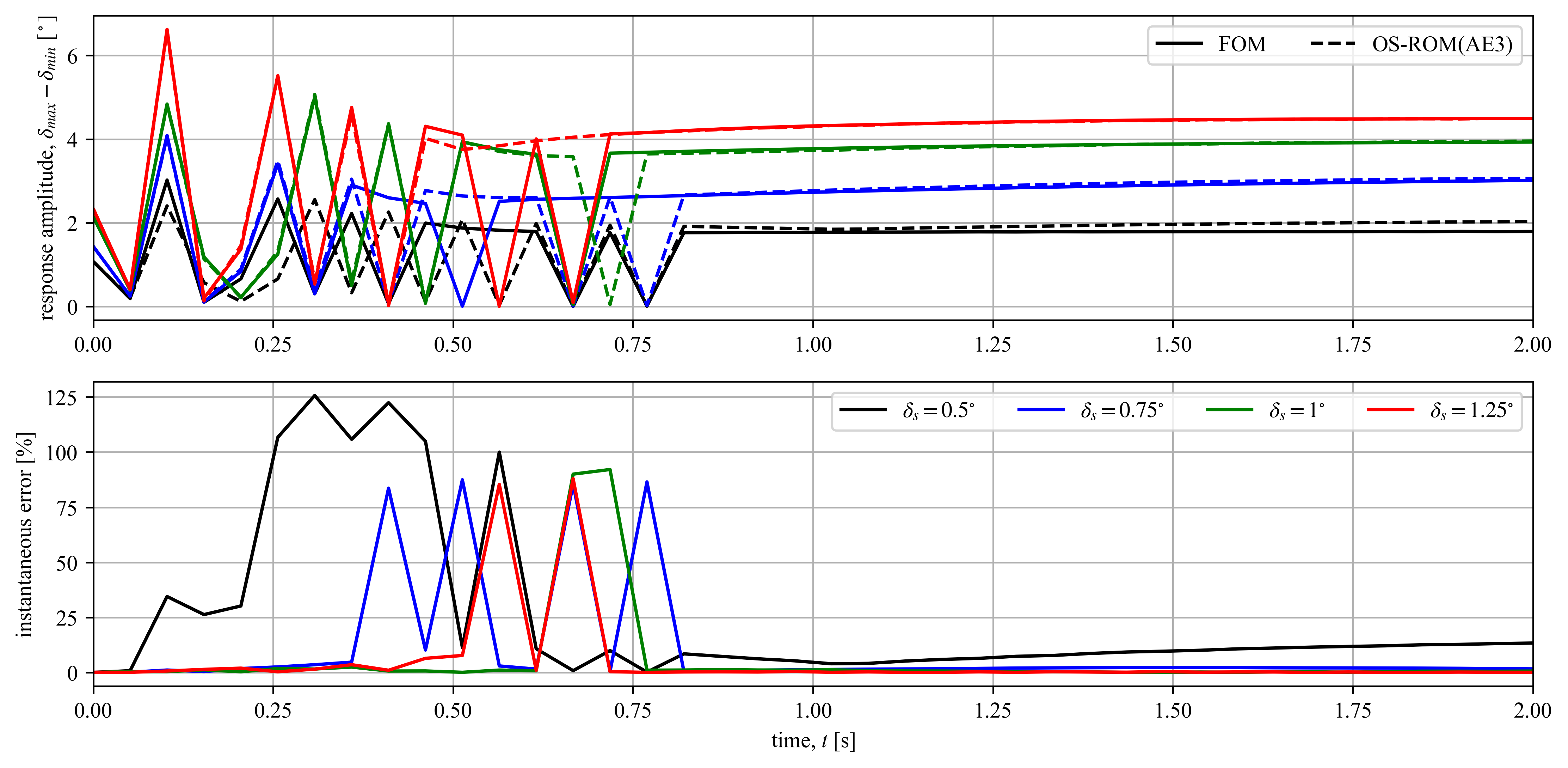}
		\caption{Time-accurate error in the response amplitude for different freeplay values comparing the FOM and OS-ROM(AE3) solutions}
		\label{fig:time_march_fp1}
	\end{figure}

   The time-marching error is now investigated for the third-order OS-ROM(AE3) as the aeroelastic response passes through the initial transient and into a bounded limit cycle. Given that some (even very small) discrepancy in the prediction of the frequency leads to phase errors, the time-accurate error is computed for the amplitude only as is shown in Fig.~\ref{fig:time_march_fp1}. For all cases the error is initially small where the response is dominated by the initial perturbation. The error then increases as the response progresses through the initial transient. As the system approaches LCO the error decays to a constant value of less than 2\% for all cases aside from $\delta_s = \pm 0.5^\circ$ (discussed above).

    \clearpage

    \subsection{Aeroelastic Response for Varying Velocity Index}
    A more rigorous test of the generalizability of the ROMs is to vary the velocity index. In the linear regime, linearized generalized aerodynamic forces scale linearly with dynamic pressure. This relationship, however, is with the assumption of small displacements and linear aerodynamic forces, therefore is not guaranteed for nonlinear transonic aerodynamic problems. In this section the velocity index is varied to assess if the nonlinear ROM will conform to this relationship. The operating conditions for which the aeroelastic optimization occurs are consistent with the previous section and denoted by $\star$.
    
\begin{figure}[h]
		\centering
            \subfigure[]{\label{m096}
			\includegraphics[width=0.47\textwidth]{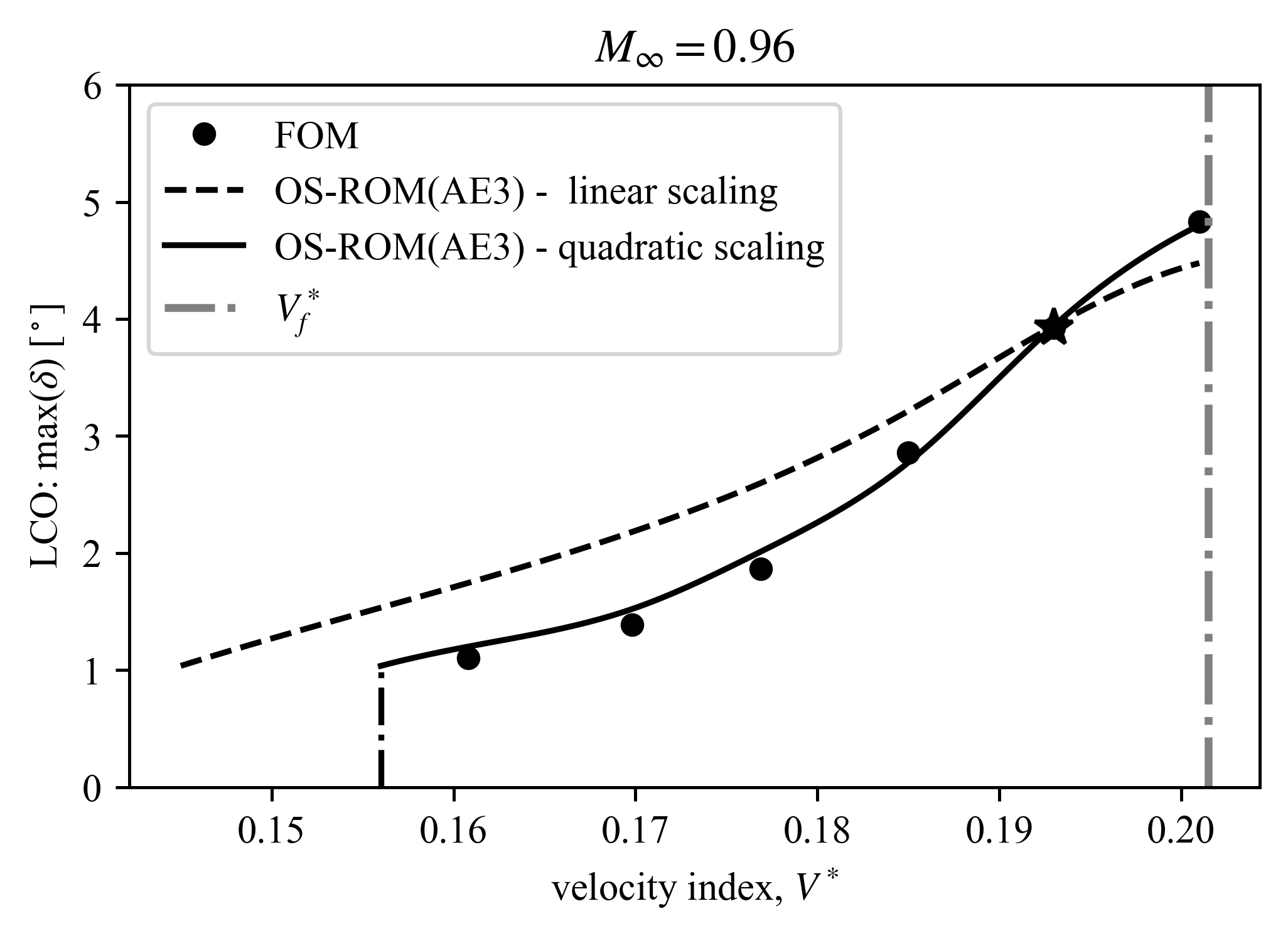}}
            \subfigure[]{\label{m0901}
			\includegraphics[width=0.47\textwidth]{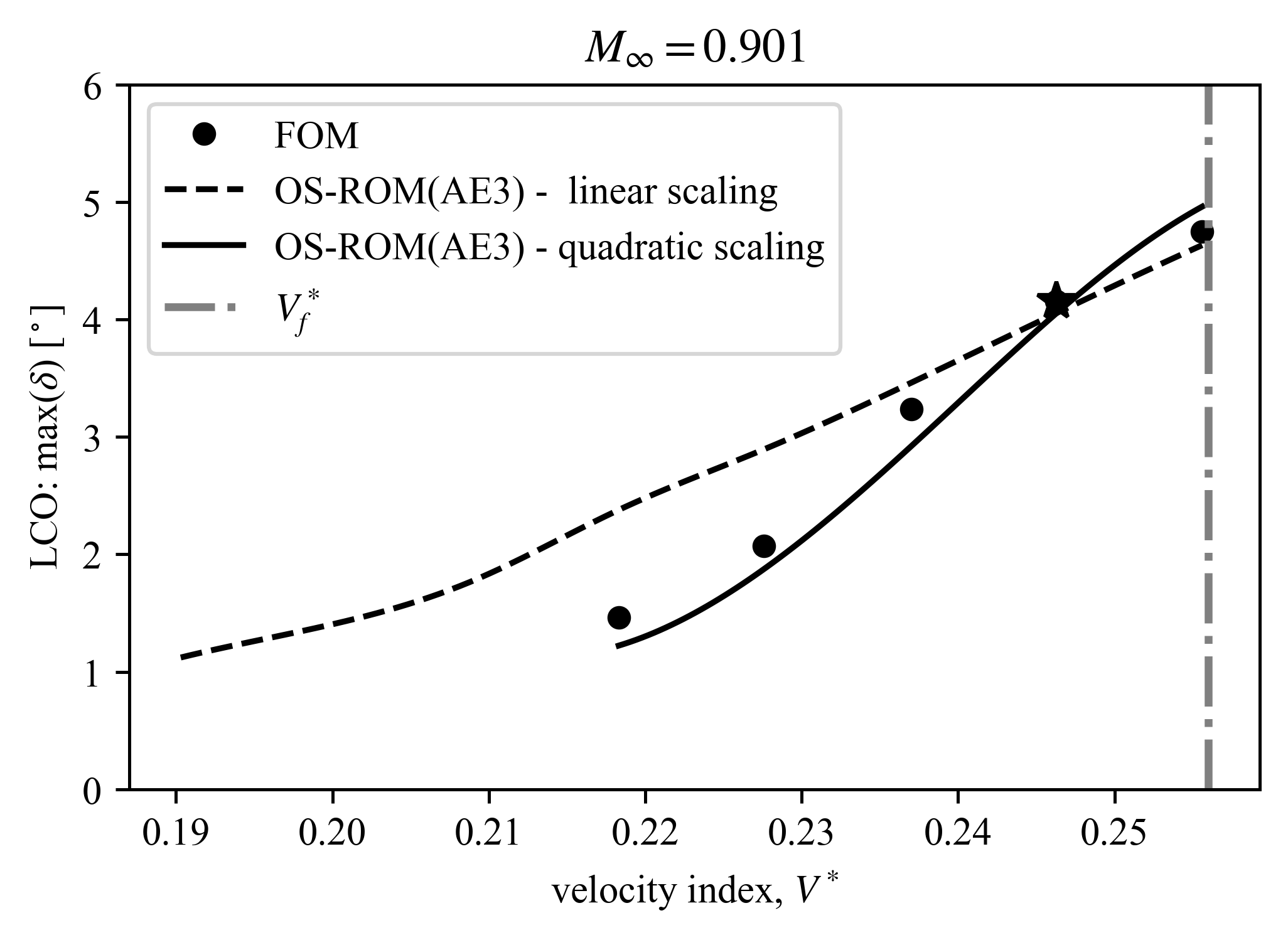}}
		\caption{LCO amplitude comparing the FOM and OS-ROM(AE3) solutions for $\delta_s = \pm 1^\circ$}
		\label{fig:q_sweep_fp1}
    \end{figure}

    It is found that while the linear relationship does not hold particularly well (although it is reasonable given the high transonic Mach number and presence of freeplay), there is a strong quadratic relationship between the generalized forces and dynamic pressure, as is presented in Fig.~\ref{m096}. A sparse third-order OS-ROM(AE3) is also identified for $M_\infty = 0.901$ (Fig.~\ref{m0901}) where it can be seen that the the linear scaling performs well for the higher velocity index values (above $V^* =0.2375$), and severely over predicts below this. On the other hand, the quadratic scaling is able to provide a better overall representation of the limit cycle amplitudes.

\begin{figure}[h]
	\centering
	\includegraphics[width=1\textwidth]{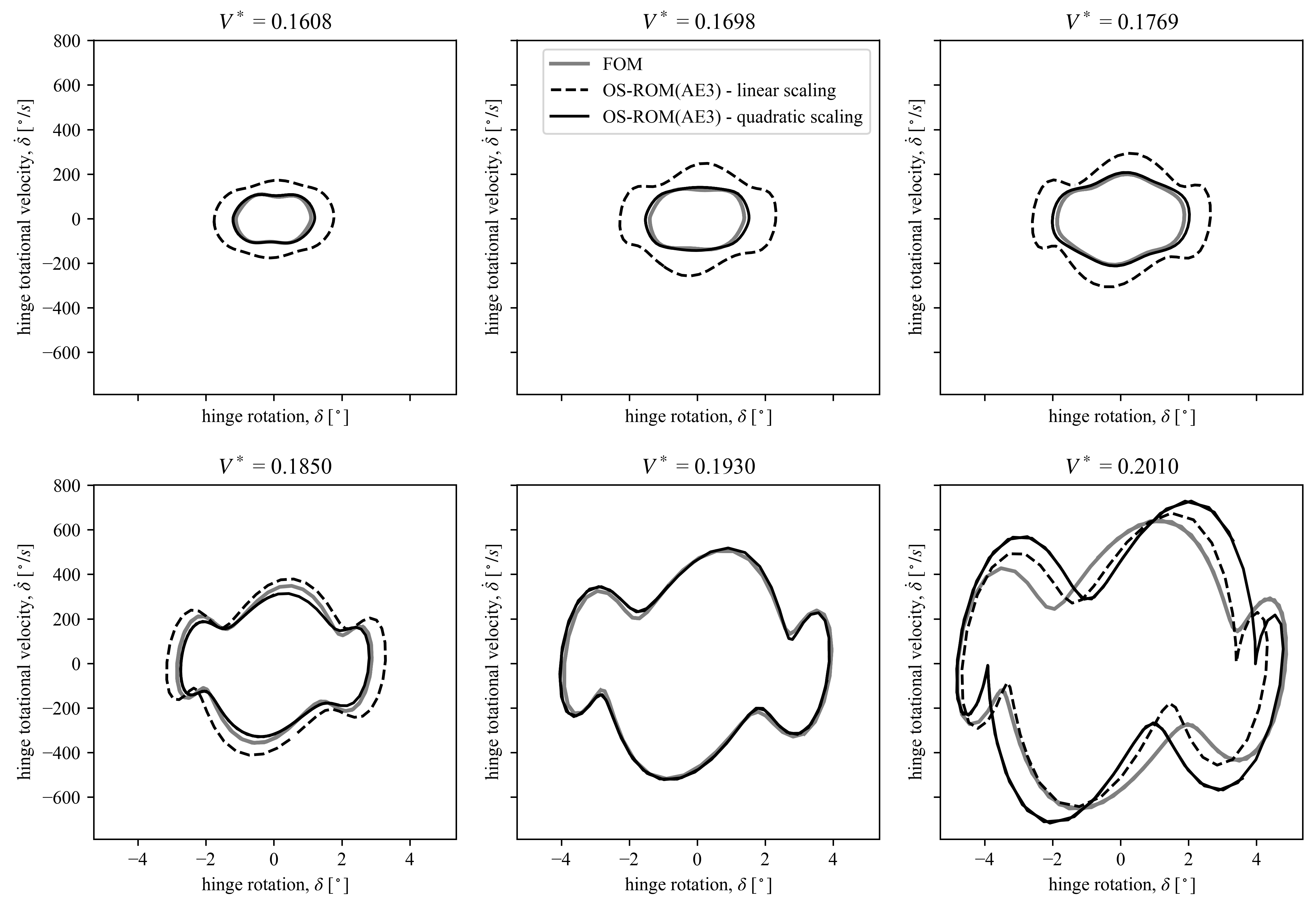}
	\caption{Phase portraits comparing the FOM and OS-ROM(AE3) solutions for $\delta_s =\pm  1^\circ$ and $M_\infty = 0.96$}
	\label{fig:q_phase}
    \end{figure}

    The phase portraits are presented in Fig.~\ref{fig:q_phase}, comparing linear to quadratic scaling of the generalized forces of the third-order ROM(AE3) with dynamic pressure for $\delta_s = \pm 1^\circ$ and $M_\infty = 0.96$. Under linear scaling, it can be seen for the three lowest velocity index values that there is an over prediction of the LCO and the ROMs do not capture the form of the LCO particularly well. This discrepancy is reduced for $V^* = 0.185$ with a smaller over prediction and the form is captured with reasonable accuracy. At $V^* = 0.201$, a moderate under prediction is observed. For this speed the form is captured poorly with a strong inflection in the asymmetrical LCO for the positive portion of the cycle that does not exist in FOM response. Under quadratic scaling the third-order OS-ROM(AE3) captures the amplitude and form very well up to $V^* = 0.193$. At $V^* = 0.201$, although the amplitude is captured with excellent precision, a small knot can be observed (an exaggeration of the linear scaling behaviour) which does not exist in the FOM response. 

\begin{figure}[h]
		\centering
			\includegraphics[width=1\textwidth]{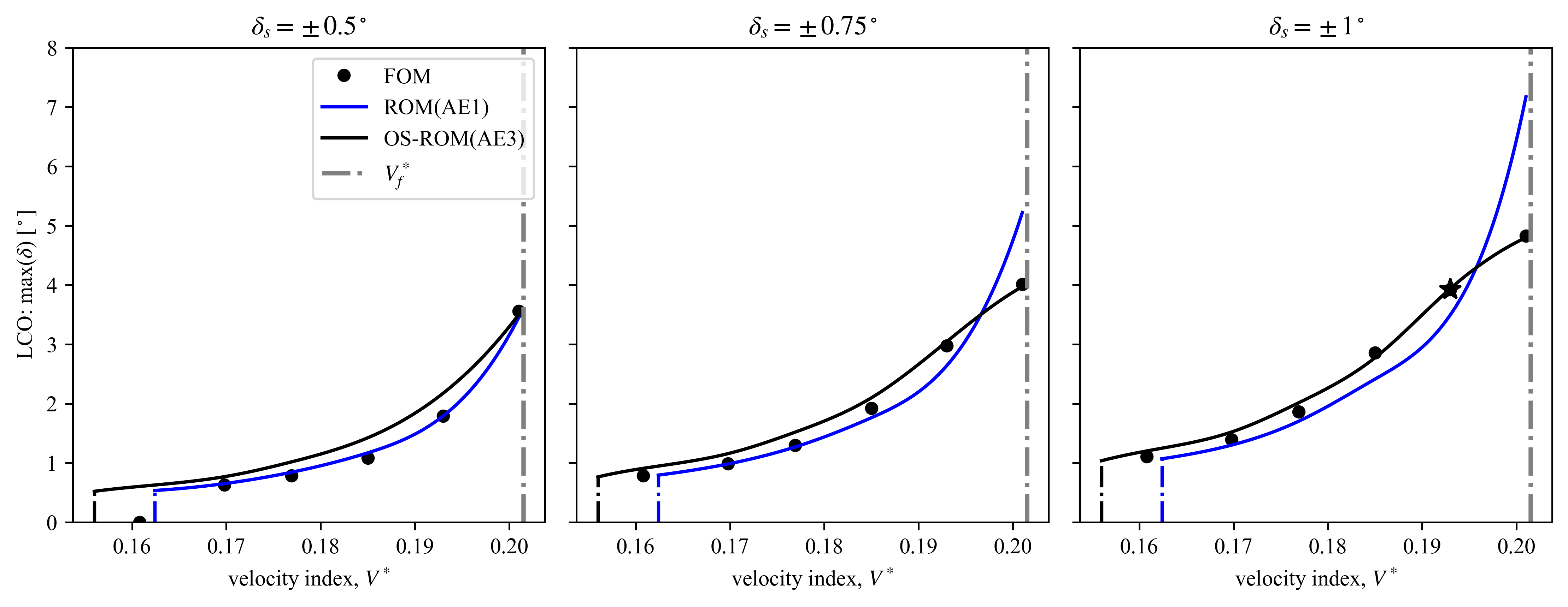}
		\caption{LCO amplitude comparing the FOM, ROM(AE1) and OS-ROM(AE3) solutions with quadratic scaling of the generalized forces for $M_\infty = 0.96$}
		\label{fig:q_sweep}
	\end{figure}

    To investigate this relationship further and ensure that it is not anomalous, the relationship is investigated for different ROMs and freeplay values, presented in Fig.~\ref{fig:q_sweep}. Under quadratic scaling, the linear ROM(AE1) performs well for the lowest freeplay value ($\delta_s =\pm  0.5^\circ$), predicting the nonlinear flutter point and the entire LCO envelope with excellent precision. For this same case, the nonlinear ROM performs well at the flutter boundary, however, slightly over predicts the LCO amplitude for lower velocity index values and under predicts the nonlinear flutter boundary. For the larger freeplay values ($\delta_s = \pm 0.75^\circ$ and $\delta_s = \pm 1^\circ$), the linear ROM(AE1) performs well for lower values of velocity index, then starts to deviate from the FOM, under predicting the LCO amplitude for values above $V^* = 0.18$. Above $V^* = 0.192$, the linear ROM(AE1) predicts exponential growth as the system approaches flutter (consistent with the expected behaviour of a linear model). On the other hand, the sparse third-order OS-ROM(AE3), is able to capture the entire LCO envelope with good precision. This includes the aerodynamic damping effects which are expected in high amplitude transonic LCO, limiting the amplitude of the response as the system approaches flutter. While the authors are not suggesting that quadratic scaling can be applied as a general rule, it is sufficiently interesting to present and warrants further investigation.

    \clearpage


    \subsection{Computational Savings}
    
        Computational performance is assessed using clock-time (actual simulation time) and CPU-h = clock-time $\times$ number-of-cores. To ensure a stable limit cycle is achieved 3000 numerical time-steps are required. Any CFD-based aerodynamic solutions are run on an Intel Xeon Gold 6152 CPU using 16-cores, while any ROM solutions on a single-core. Table~\ref{tab:compsav} presents the computational cost associated with the various solution strategies presented in this paper. In terms of simulation time, the ROM solutions are several orders of magnitude faster than the FOM. A further benefit of using a sparse representation can be observed in the computational time for full third-order ROM versus sparse third-order ROM solutions, with the sparse ROM solutions being two-three times faster. The time taken to generate the ROM must also be taken into account. The aerodynamic-optimized ROM takes approximately 32 CPU-h ($\sim$2 hours clock-time) to generate, while the aeroelastic optimized ROM takes approximately 100 CPU-h ($\sim$8 hours clock-time)to generate (as it requires a FOM aeroelastic solution as the objective function. See the following section for further discussion.

        \clearpage

\begin{singlespace}
\begin{table}[h]
    \centering
    \begin{tabular}{llccc}
        \hline
         Task & Details & Clock-time &  CPU-h & Savings \\
        \hline
        \hline
        \textbf{Offline Costs}& \\
        \hline
        Aerodynamic-Optimized ROM & \\
        \hline
        Generate Aerodynamic Data & CFD run with random excitation & 2 & 32 \\
        Grid search & Hyperparameter tuning (AD) & 0.1-0.5 & 0.1-0.5 \\
        \hline
        &  \multicolumn{1}{r}{Total:} &2.1-2.5& 32.1-32.5 \\
        \hline
        Aeroelastic-Optimized ROM & & \\
        \hline
        Generate Aerodynamic Data & CFD run with random excitation & 2 & 32 \\
        Generate Aeroelastic Data & One FOM run for 1000 time-steps & 4 & 64\\
        Grid search & Hyperparameter tuning  (AE) & 2-4& 2-4\\
        \hline
        &  \multicolumn{1}{r}{Total:} & 8-10 & 98-100 \\
        \hline
        \textbf{Online Costs} & & \\
        \hline
         FOM solution & One run for 3000 time-steps &12 & 192  \\
         ROM (linear) solution & One run for 3000 time-steps & 0.0125& 0.0125& 15360$\times$  \\
         ROM (third-order) solution & One run for 3000 time-steps & 0.166& 0.166& 1152 $\times$  \\
         D-ROM solution & One run for 3000 time-steps& 0.066 & 0.066 & 2880$\times$ \\
         L-ROM solution & One run for 3000 time-steps & 0.066& 0.066 & 2304$\times$ \\
         OS-ROM solution & One run for 3000 time-steps & 0.05 & 0.05 & 3840$\times$ \\
         \hline 
    \end{tabular}
    \caption{Computational cost and savings for the various ROM methodologies}
    \label{tab:compsav}
\end{table} 
\end{singlespace}

    \section{Discussion}
    \label{sec:disc}

    This section aims to provide the reader with a detailed discussion surrounding the limitations of the proposed approach and opportunities to progress this research, summarized as follows: 
    \begin{itemize}
    \item \textbf{Multi-input identification:} Although in this paper nonlinear cross-terms are neglected entirely it should not be considered as a generalization of this approach. The need to include nonlinear cross-terms varies on a case-by-case basis and is often shown to significantly improve the accuracy of the ROM. An opportunity to extend would employ sparsity promotion to identify the multi-input system. The use of an algorithm like OMP becomes even more crucial for the multi-input problem given that the number of terms also increases exponentially with the number of structural modes which becomes a second avenue for the curse-of-dimensionality.
    \item \textbf{Real-world problems:} Although this work makes a significant step forward toward applying this class of ROM to real-world aeroelastic problems the extension to a full aircraft model still comes with a significant computational burden considering that $i$) nonlinear generalized forces need to be identified for dozens of structural modes, and $ii$) the CFD model would generally be at least an order of magnitude larger. For multi-input identification of a full aircraft, the number of nonlinear terms (including cross-terms) without sparsity to be identified would be in the order of hundreds-of-millions or billions. Significant opportunities exist in extension of this class of nonlinear ROM with sparsity promotion to full aircraft models.
    \item \textbf{Offline computational cost:} The ROM use case should be justified before deciding whether to use the aerodynamic-optimized or aeroelastic-optimized variants of the OS-ROM. If the ROM is to be used online a large number of times then the added offline computational cost (and slightly improved accuracy) may be justified. Otherwise, it may be preferable to optimize for aerodynamic performance with knowledge of a slight decrease in aeroelastic performance which allows the ROM to be generated approximately 4-5$\times$ faster.
    
    \end{itemize}

    \section{Summary and Conclusion}
    \label{sec:conc}
    A new approach to identify nonlinear aeroelastic ROMs is presented, based on automatic identification of optimal sparsity patterns in the Taylor partial derivatives of the unsteady aerodynamic forces. Several sparsity inducing approaches are implemented, including, Orthogonal Matching Pursuit, LASSO regression and rigid sparsity selection. The findings highlight that it is preferable to have control over the $\ell_0$ pseudo-norm than the $\ell_1$-norm for this class of problem, while using a rigid sparsity definition is prone to over fitting.
    
    Through OMP, which is shown to outperform the other methods in terms of sparsity, accuracy and generalizability, it is possible to rapidly identify the optimal $s$-sparse coefficients of the higher-order Taylor partial derivatives for efficient and accurate nonlinear aeroelastic modeling. The case study is an all-movable horizontal tail model with freeplay undergoing high-amplitude limit cycles at zero-AoA and high transonic Mach numbers. By estimating less than 20 of 500+ terms, from the first-order, second-order, and third-order partial derivative tensors, the optimized sparse ROM is able to model the nonlinear transonic aeroelastic LCOs with excellent precision, and generalize to new freeplay values, with online computational savings of several orders of magnitude. In terms of the ROM performance for new velocity index values, the linear relationship between generalized force and dynamic pressure is weak for this complex nonlinear problem. However, it is shown that accurate prediction of the LCO amplitude can be obtained through nonlinear scaling of the generalized forces with dynamic pressure--a prospect worthy of ongoing investigation. Follow up articles will focus on addressing sensitivities to parameter changes in a higher-dimensional parameter space and multi-input identification using OMP. 

    \section*{Acknowledgments}
The authors are grateful for the ongoing financial support provided by the Australian Defence Science and Technology Group (DSTG).
        \appendix
        \section{Hyperparameter Grid Search Results}
        \label{app:gridsearch}
\begin{figure}[h]
		\centering
		\subfigure[]{\label{mode1_free}
			\includegraphics[width=0.45\textwidth]{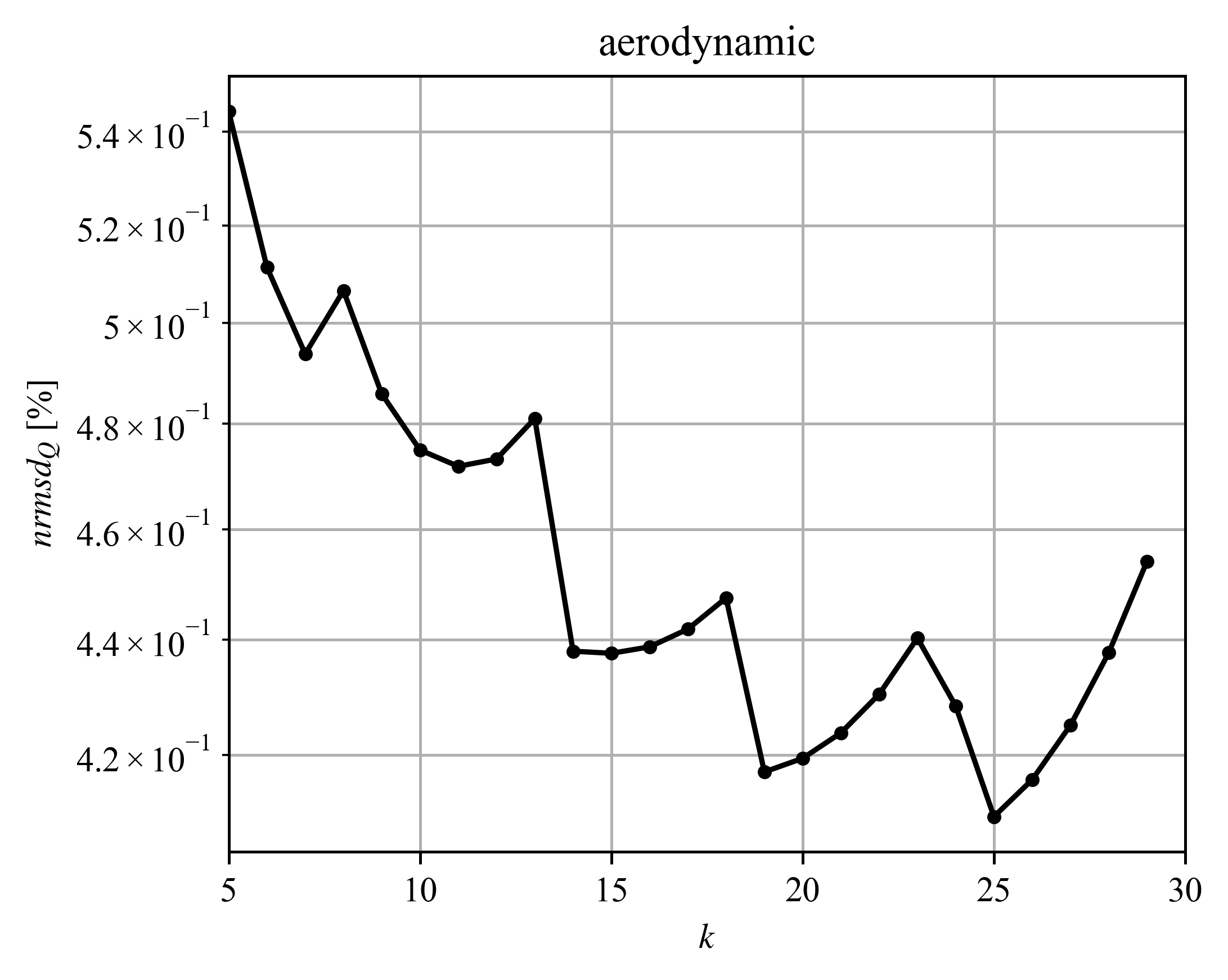}}
		\subfigure[]{\label{mode2_free}
			\includegraphics[width=0.45\textwidth]{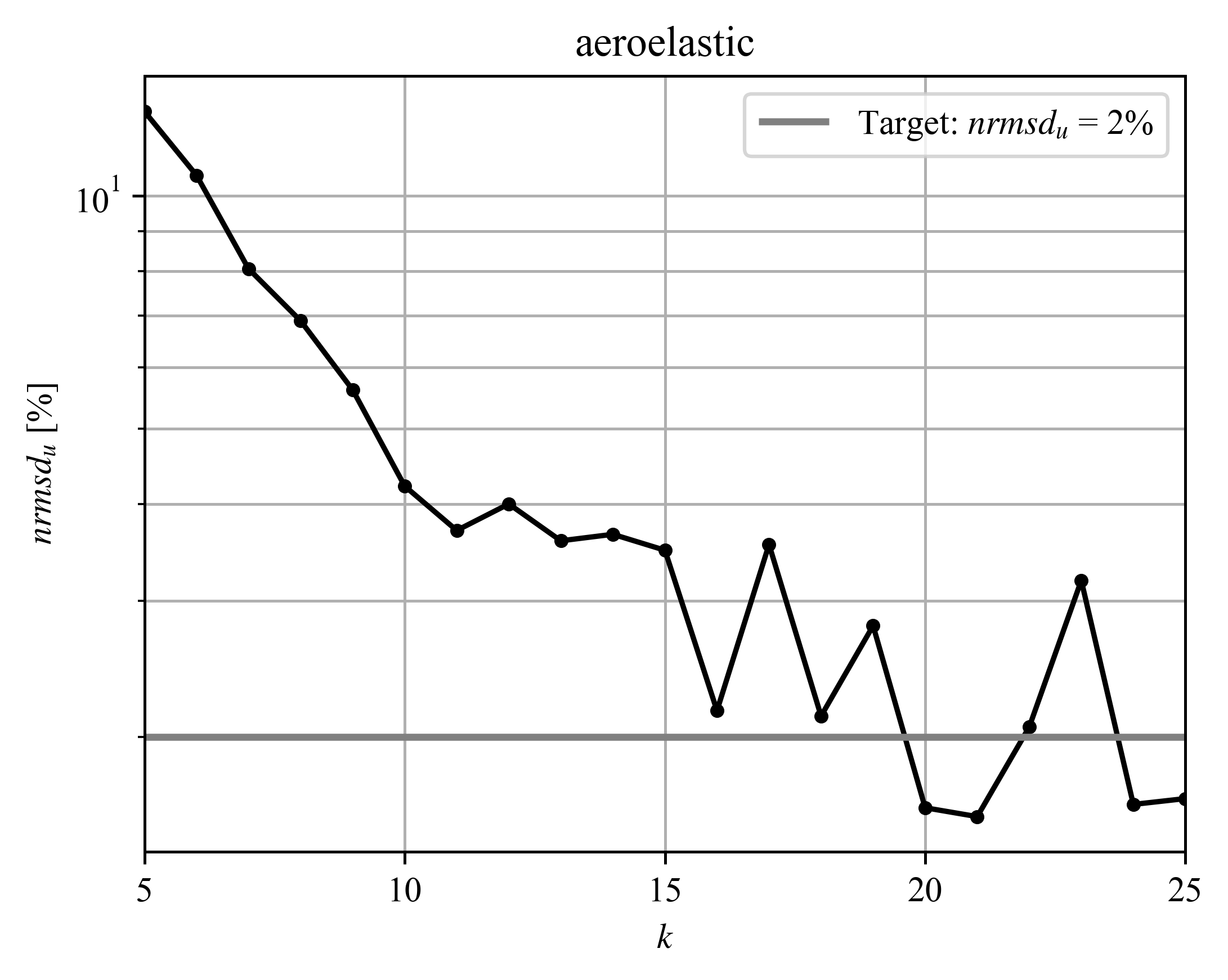}}
		\caption{D-ROM grid search results; a) aerodynamic response and b) aeroelastic response}
		\label{fig:d_gs}
	\end{figure}

        The target of the aeroelastic grid search is $nrmsd_\delta < 2\%$. Figure~\ref{fig:d_gs} present the results for the D-ROM where it can be seen that optimal aerodynamic performance is achieved with 25 lag terms, while for the aeroelastic response it is with 16 lag terms. There is reasonable correlation between the two curves where the performance is good for the region greater than 15 lag terms. 

\begin{figure}[h]
		\centering
		\subfigure[]{\label{mode1_free}
			\includegraphics[width=0.375\textwidth]{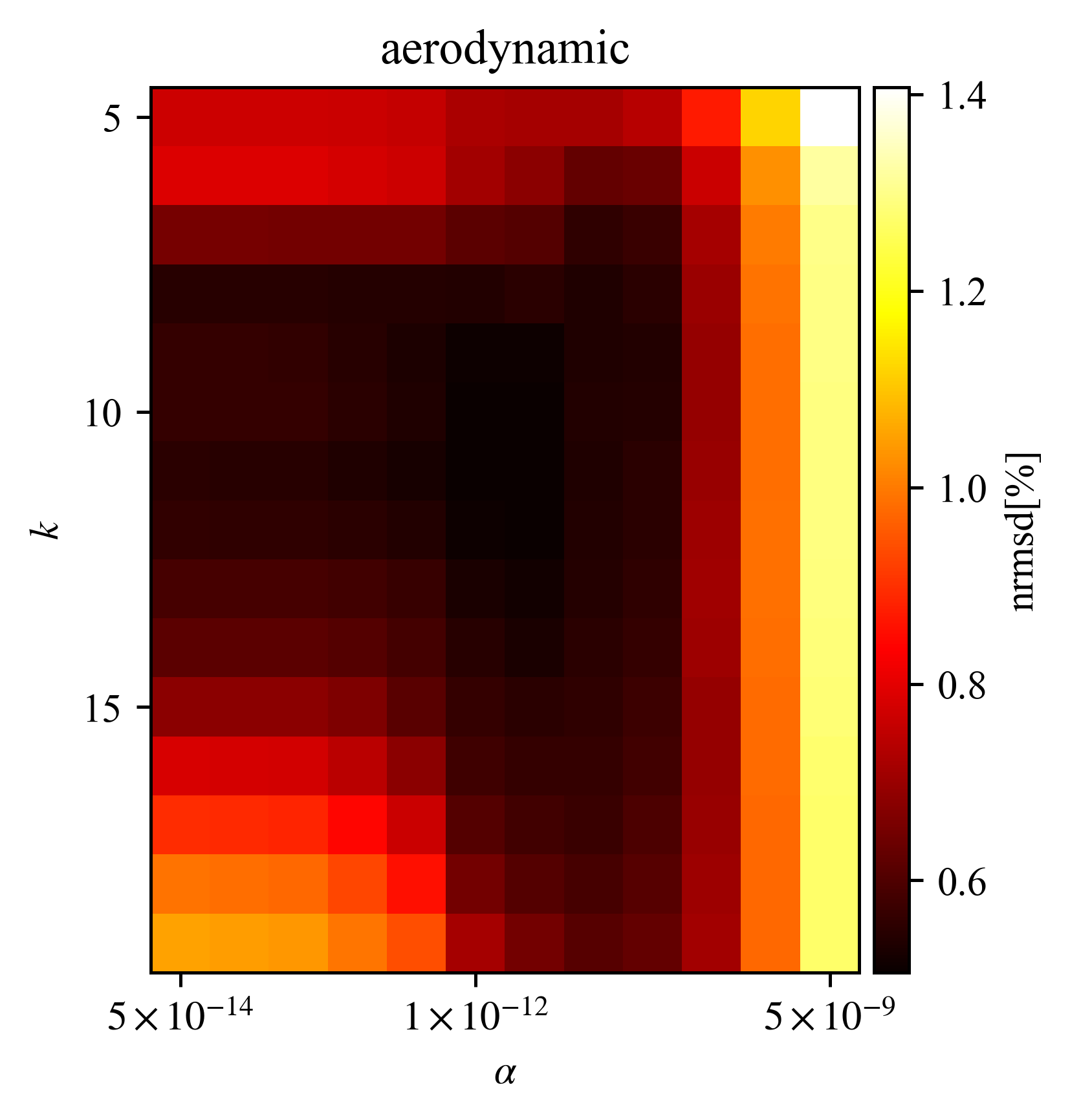}}
		\subfigure[]{\label{mode2_free}
			\includegraphics[width=0.375\textwidth]{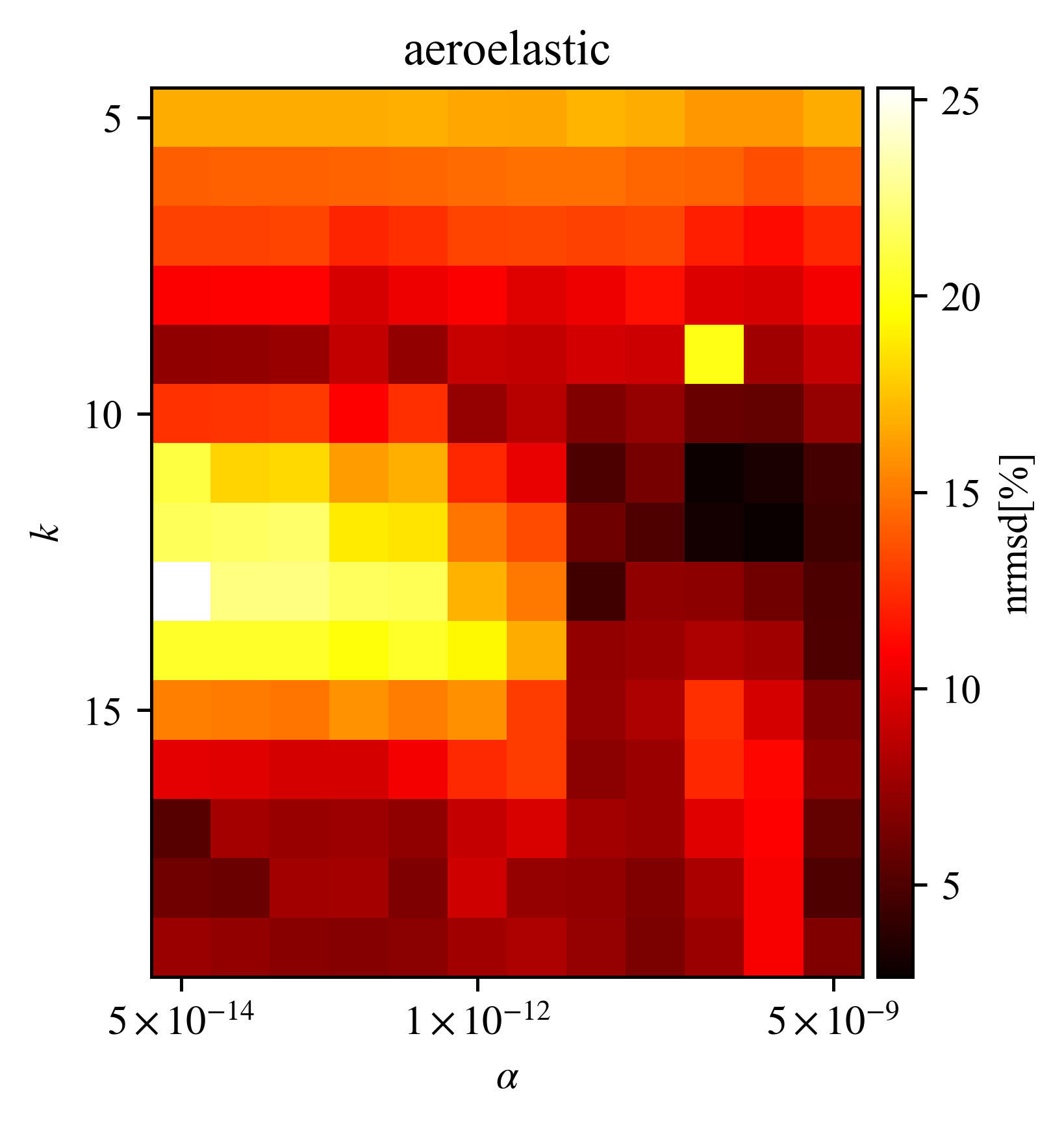}}
		\caption{L-ROM grid search results; a) aerodynamic response and b) aeroelastic response}
		\label{fig:l_gs}
	\end{figure}

         Figure~\ref{fig:l_gs} presents the aerodynamic and aeroelastic grid search results for the L-ROM where the heat map demonstrates very little correlation between the two optimization problems, $i.e.$, the global minima occur in quite different regions of the parameter space. Furthermore, the target $nrmsd_\delta < 2\%$ is only achieved in a very small region of the hyperparameter space for the aeroelastic optimization problem. 

\begin{figure}[h]
		\centering
		\subfigure[]{\label{mode1_free}
			\includegraphics[width=0.45\textwidth]{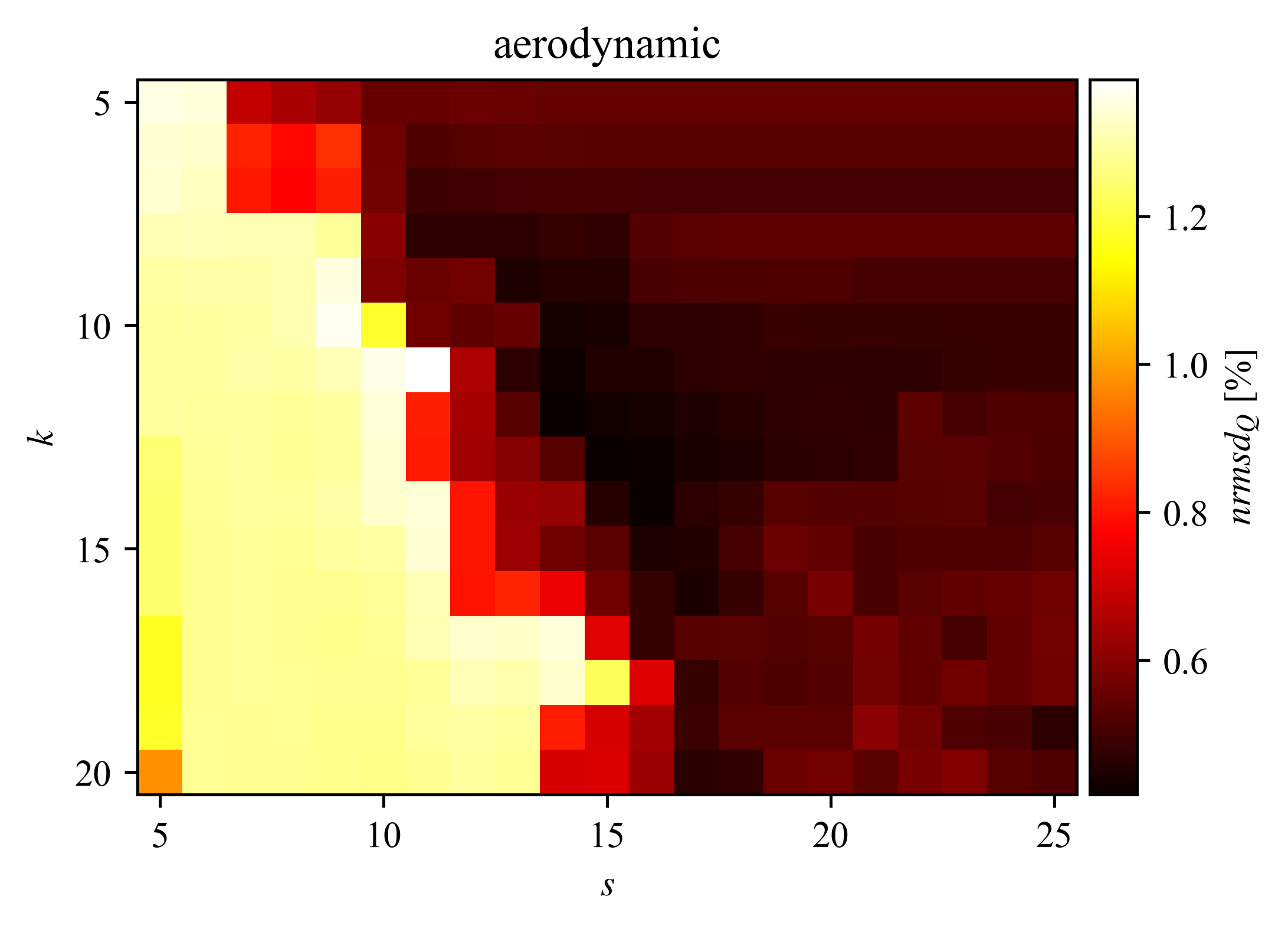}}
		\subfigure[]{\label{mode2_free}
			\includegraphics[width=0.45\textwidth]{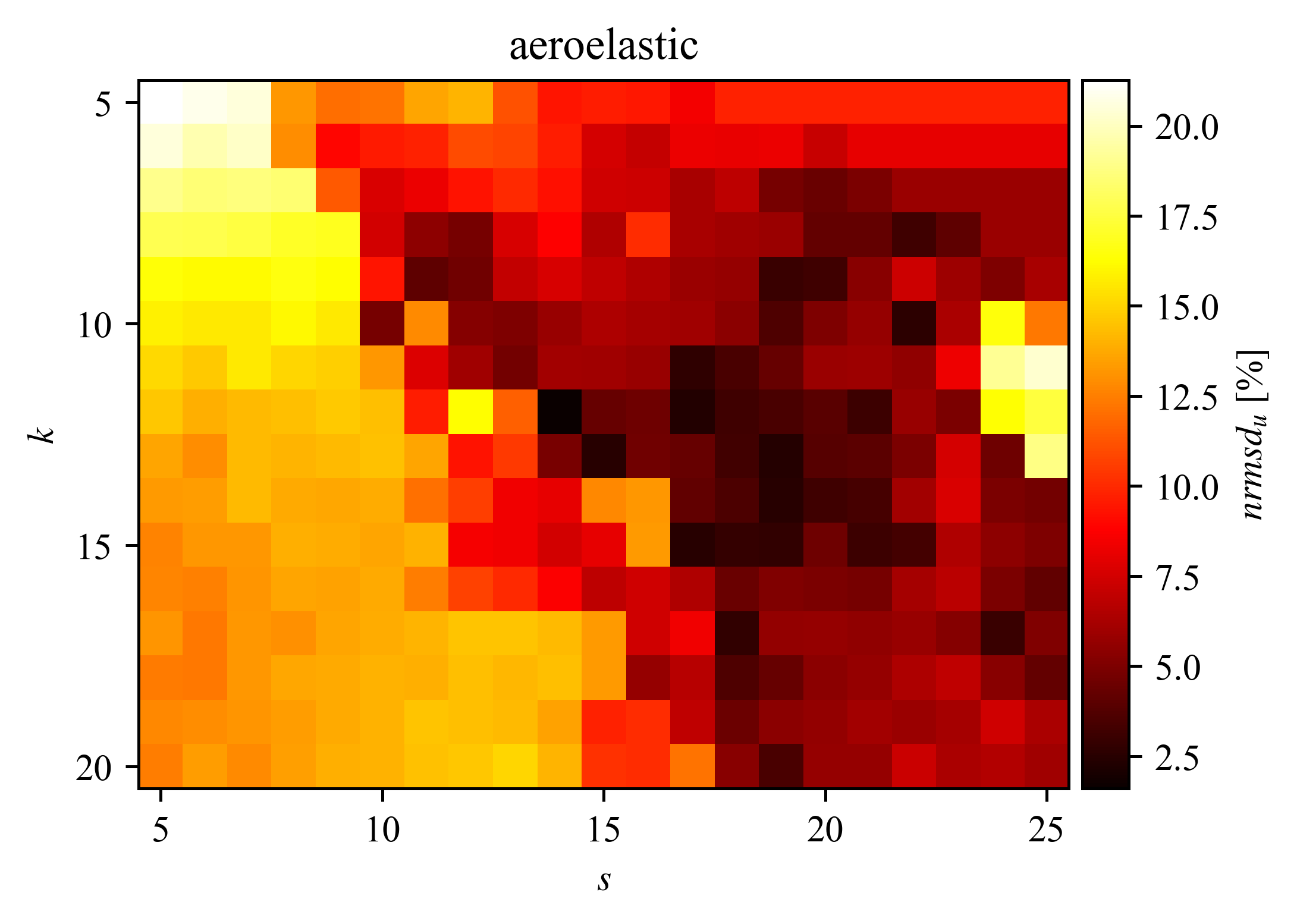}}
		\caption{OS-ROM grid search results; a) aerodynamic response and b) aeroelastic response}
		\label{fig:omp_gs}
	\end{figure}

         Figure~\ref{fig:omp_gs} presents the aerodynamic and aeroelastic grid search results for the OS-ROM where good correlation between the two optimization problems can be observed. Most notably, the global minimum occurs in a consistent region of the hyperparameter space for both.  


        
        \clearpage
        \bibliographystyle{IEEEtranDOI}
        \bibliography{gensys_doi_2}

\end{document}